\documentstyle[prc,preprint,tighten,aps]{revtex}
\begin {document}
\draft
\title{Structure of the mirror nuclei $^9$Be and $^9$B in a microscopic
cluster model}
\author{K. Arai$^{1}$, Y. Ogawa$^{2}$, Y. Suzuki$^{3}$, and K. Varga$^{4}$ 
\
\
\\ $^1$ Graduate School of Science and Technology, Niigata University, 
Niigata 950-21, Japan
\\ $^2$ RIKEN, Hirosawa, Wako, Saitama 351-01 Japan
\\ $^3$ Department of Physics, Niigata University, Niigata 950-21, Japan
\\ $^4$ Institute of Nuclear Research of the Hungarian Academy of Sciences,
\\Debrecen, P. O. Box 51, H--4001, Hungary}
\date{\today}
\maketitle
\begin{abstract}
\noindent
The structure of the mirror nuclei $^9$Be and $^9$B is studied in a 
microscopic
$\alpha+ \alpha+ n$ and $\alpha+ \alpha+ p$ three-cluster model  
using a fully antisymmetrized 9-nucleon wave function. The two-nucleon  
interaction includes central and spin-orbit components together with 
the Coulomb potential. The ground state of $^9$Be is obtained
accurately with the stochastic variational method, while several 
particle-unbound 
states of both $^9$Be and $^9$B are investigated with the 
complex scaling method. 
The calculation for $^9$Be supports the recent identification 
for the existence of two broad states around 6.5 MeV, and predicts 
the $\frac{3}{2}^{-}_2$ and $\frac{5}{2}^{-}_2$ states 
at about 4.5 MeV and 8 MeV, respectively. The 
similarity of the calculated spectra of $^9$Be and $^9$B enables one to 
identify unknown spins and parities of the $^9$B states.        
Available data on electromagnetic moments and elastic electron 
scatterings are reproduced very well. The enhancement of the $E$1 transition 
of the first excited state in $^9$Be is well accounted for. 
The calculated density of $^9$Be 
is found to reproduce the reaction cross section on a Carbon target. 
The analysis of the beta decay of $^9$Li to $^9$Be   
clearly shows that the wave function of $^9$Be must contain a small 
component that cannot be described by the simple $\alpha+ \alpha+ n$ 
model. This small component can be well accounted for 
by extending a configuration space to include the distortion of the  
$\alpha$-particle to $t+p$ and $h+n$ partitions.

\end{abstract}


\narrowtext

\section{Introduction}
There has been a growing interest in the study of neutron-rich nuclei since
 the advent of radioactive nuclear beams. It was found \cite{tanihata} that 
some light nuclei near the neutron drip-line exhibit neutron-halo structure 
or have thick neutron-skin clouds. 
The halo structure, a new form of the nuclear matter, is 
characterized by a spatially extended low density distribution around the 
core part of  
normal density. It is interesting to know how a nucleus changes its 
structure with the increase of the number of neutrons  and 
how the binding of the neutrons is attained in such a system.  
In the very light nuclei the mean field is not stable enough to 
generate the regular shell structure but, instead, the clustering of the 
nucleons, especially the $\alpha$-clustering plays an important role in 
determining their structure. 
Because of this the light nuclei show individual features which have 
strong dependence on the number of nucleons. The Be isotopes are of 
special interest in this respect because they show  some anomalous 
features which are not easily understood in a simple shell model. 
Because the $^8$Be nucleus is known to be a typical cluster state 
of two $\alpha$-particles, it is interesting to attempt at describing  
heavier Be isotopes in a unified framework of two $\alpha$-particles and  
extra neutrons. Our basic question is: How well does this 
picture give us a consistent understanding of the Be isotopes?  This 
question  
naturally leads us to the application of a multicluster model. A fully  
microscopic multicluster model utilizes an A-nucleon wave function, 
incorporating the Pauli principle exactly. It has various  
applications in the structure study for the halo nuclei \cite{suzuki} 
and in the nuclear astrophysics \cite{langanke,baye}.  
\par\indent
The spectrum of $^9$Be is poorly known. This is probably because all the 
levels but the ground state are above the $\alpha+ \alpha+ n$ threshold. 
Recent experiments \cite{glickman,dixit} have, however, isolated the 
broad level at 
6.76 MeV \cite{ajzen} to two states, the $\frac{7}{2}^-$, 6.38 MeV state 
and the $\frac{9}{2}^+$, 6.76 MeV state.
\par\indent
A few theoretical studies on $^9$Be have already been done in various 
models. 
A projected Hartree-Fock calculation \cite{bouten} was carried out 
to study the 
electromagnetic properties of $^9$Be. A shell-model calculations in 
a (0+1)$\hbar \omega$ basis \cite{glickman,hees84} gave a reasonable 
spectrum but predicted too small dipole transition strength for the 
first excited state. 
There are several calculations using an $\alpha+\alpha+N$ 
three-cluster model. Earlier 
calculations \cite{fonseca,orlo} emphasized the three-body aspect 
of $^9$Be to explain its low-lying spectrum. These treated the 
$\alpha$-particle as a structureless particle and considered its 
compositeness by redefining the potential with the Pauli correction. 
Recently, this type of macroscopic approach has been extensively applied 
to the study of $^9$Be and $^9$B nuclei \cite{voron}, by including the    
$(\alpha \alpha)N$-type arrangement in the calculation. 
On the other hand, some microscopic cluster-model 
calculations starting from  9-nucleon wave functions were accomplished 
in the resonating group method \cite{zahn} or in the 
generator coordinate method \cite{okabe,suppl,desc}. Our microscopic 
multicluster 
model has the advantage that the distortion of the constituent clusters, 
e.g., the $\alpha$-particle, when needed, can be included in the 
calculation in a consistent way. An example indicating this necessity will 
be discussed later in  
case of the $\beta$ decay of $^9$Li to $^9$Be. The macroscopic model has,    
however, a difficulty in taking the possibility of the cluster distortion 
into account.  
\par\indent
The calculation of Ref. \cite{zahn} 
considered the three channels of $^8$Be(0$^+$)+$n$, $^5$He(
g.s.)+$\alpha$, and $^8$Be(2$^+$)+$n$ to describe the levels of $^9$Be. A 
molecular model was applied in the generator coordinate framework to study 
the structure of $^9$Be \cite{okabe,suppl}. The calculation of Ref. 
\cite{desc} included only $^8$Be+$N$ channel, where the motion of the two 
$\alpha$-particles in $^8$Be was described in a restricted space. The two 
generator coordinate method calculations gave a reasonable agreement 
with experiment. There are, however, some noticeable disagreements between 
the theory and experiment. Both of the macroscopic and microscopic 
calculations done so far were limited either in taking   
the Pauli principle into account or in  treating the three-body dynamics.  
Further improvement will be attainable by treating the 
three-body dynamics more completely.  
\par\indent
As the first of the series of studies on the Be isotopes we show in this 
paper the results of calculation for $^9$Be in a microscopic 
$\alpha+\alpha+n$ 
model. At the same time we consider its mirror nucleus $^9$B in 
an $\alpha+\alpha+p$ three-cluster model. One of the main objectives in this 
paper is to assess the validity of our basic assumption in case of $^9$Be. 
This is substantially important for the study on heavier Be isotopes. 
To this end we carry out an extensive 
three-cluster model calculation that has no limitations mentioned above, 
and investigate carefully some 
important properties of the low-lying states of $^9$Be, that is, the energy  
spectrum, the magnetic and quadrupole moments of the $\frac{3}{2}^-$ ground 
state, 
and the electron scattering form factors. Of particular interest is the 
enhancement of the 
electric dipole transition from the first excited $\frac{1}{2}^+$ state to 
the ground state \cite{millener,barker}. 
This reduced transition probability is nearly  as large as the well known 
one of $^{11}$Be. The mechanism of the enhancement in these cases may be 
related to each other. Another interest is the $\beta$ decay from $^9$Li to 
the low-lying states of $^9$Be \cite{nyman}. 
We will show that this $\beta$ decay is useful to reveal small components 
contained in the wave function of $^9$Be. 
\par\indent
The levels of $^9$B are all particle-unbound and only few of them have spin 
assignments \cite{ajzen}. 
There are discussions on the missing $\frac{1}{2}^+$  
state from the viewpoint of the Coulomb displacement energy 
\cite{sherr,tiede}. As a mirror nucleus of $^9$Be, 
$^9$B can be described in an $\alpha+\alpha+p$ three-cluster model. 
A cluster model has a unique advantage that it can describe the asymptotic 
part of a wave function well and thereby predict the position and 
width of a resonance. This is a very important ingredient for a detailed 
structure study of both $^9$Be and $^9$B because their  
states are mostly unbound. 
\par\indent    
In our  approach  the total wave function is given as 
an antisymmetrized product of the internal states of the clusters 
and the function of the
relative motion. The antisymmetrization of all the nucleons is exactly  
taken into account. Two types of cluster arrangements, $(\alpha \alpha)N$ 
and $(\alpha N) \alpha$, are combined to include
the different correlation between the clusters. The nucleon in 
the $(\alpha \alpha)N$ arrangement corresponds to moving in a ``molecular'' 
orbit around 
the $^8$Be=$(\alpha \alpha)$ core. On the contrary, the $(\alpha N) \alpha$ 
arrangement is suited to describe an ``atomic'' orbit of the nucleon around 
the $\alpha$-particle. This  analogy should not, however, be taken so 
literally particularly when the particles come closer, because the 
configurations of the two arrangements have considerable overlap. 
The function of the relative motion is approximated by a linear combination 
of nodeless harmonic-oscillator
functions of different size parameters. Our experience \cite{VST,VS} shows 
that the approximation with such functions gives an accurate 
description up to large distances. To keep the dimension of the basis
low, we apply the stochastic variational method (SVM) \cite{VS,Kukulin,VLS}, 
in which we set up  the ``important'' basis states stepwise by using an 
admittance test. This procedure was successfully applied to study the  
exotic nuclei \cite{suzuki,VLS,VSO} and also to few-body systems \cite{VS}. 
\par\indent
The plan of this paper is as follows. In  section II we give a brief
outline of our formalism. The microscopic three-cluster model is presented 
in subsection IIA. The scaling methods which we apply to determine
the position and width of a resonance state are briefly explained in 
subsection IIB. 
 Section III contains the results of calculations. The input parameters are 
given in subsection IIIA. The relative importance of the arrangements and 
the angular momentum channels are discussed in subsection IIIB. Energies, 
radii, magnetic and quadrupole moments, electron scattering form 
factors are compared with experiment in subsection IIIC.  The density 
distributions and the spectroscopic amplitudes are discussed in subsection 
IIID. The $\beta$ decay of $^9$Li to the states of $^9$Be is discussed in 
subsection IIIE. In the last section we summarize the most important
conclusions.

\section{FORMALISM}
\subsection{A microscopic three-cluster model}
To describe the system consisting of $\alpha+ \alpha+ n$ for $^9$Be or of 
$\alpha+ \alpha+ p$ for $^9$B, we
build up a trial function which is a sum over two cluster 
arrangements $\mu$, $\mu_1$ = $(\alpha \alpha)N$ and $\mu_2$ = $(\alpha N) 
\alpha$, with $N$ = $n$ or $p$.
Each arrangement is associated with a particular set of intercluster 
Jacobi coordinates \mbox{\boldmath $\rho$}$_1^\mu$ and 
\mbox{\boldmath $\rho$}$_2^\mu$. The coordinates \mbox{\boldmath $\rho$}$_1^\mu$ and 
\mbox{\boldmath $\rho$}$_2^\mu$ in the arrangement $\mu_1$ are chosen 
to stand for the relative 
coordinate of the $\alpha$-particles and the nucleon coordinate 
measured from the center-of-mass coordinate of two 
$\alpha$'s, while, in the arrangement $\mu_2$, they represent the relative 
distance vector between the nucleon and $\alpha$ and the relative 
coordinate of another $\alpha$ from the center-of-mass coordinate of the 
nucleon and $\alpha$. The arrangement $\mu_1$ is suited to 
describe the component corresponding to the $^8$Be+$N$ decomposition 
at large distances, while the arrangement $\mu_2$ corresponds to the 
$^5$He+$\alpha$ decomposition.  
The total orbital
angular momentum $L$ is obtained by coupling the orbital angular momenta 
 $\ell_i\equiv \ell_i^\mu$ belonging to the Jacobi coordinates
\mbox{\boldmath $\rho$}$_i^\mu$, and then it is coupled 
with the total spin $S$ = $\frac{1}{2}$ to get the total angular 
momentum $J$.  See Fig. 1(a). The intrinsic wave function of the 
$\alpha$-particle is constructed from a 
harmonic-oscillator Slater determinant with a fixed size parameter 
by eliminating the center-of-mass motion. 
The wave function of the intercluster motion
is approximated by a linear combination of nodeless harmonic-oscillator
functions (or ``Gaussians'') of different size parameters:
\begin{equation}
\Gamma_{\ell m}(\nu,\mbox{\boldmath $\rho$})=
G_{\ell}(\nu)
{\rm exp}(-\nu{\rho^2})
{\cal Y}_{\ell m}(\mbox{\boldmath $\rho$}),
\label{Gamma}
\end{equation}
with
\begin{equation}
G_{\ell}(\nu)=
\left[{2^{2\ell+7/2} \nu^{\ell+3/2}\over\sqrt{\pi}(2\ell+1)!!}
\right]^{1/2} , \ \ \ \ \ \ {\cal Y}_{\ell m}({\bf x})=x^{\ell} 
Y_{\ell m}({\hat {\bf x}})
. 
\end{equation}

\par\indent
The wave function with the angular momenta
 $[S,(\ell_1\ell_2)L]JM$ ($S$ =$\frac{1}{2}$) 
in the arrangement $\mu$ can be written as
\begin{equation}
\Psi ^{\mu}_{[S,(\ell_1\ell_2)L]JM}
=\sum_{K} C^{\mu}_{K,S(\ell_1\ell_2)L} {\cal A}\left
\{\left [\Phi _S
\left [{\Gamma }_{\ell_1}(\nu^{\mu}_{k_1},
 \mbox{\boldmath $\rho$}^{\mu}_1){\Gamma}_{\ell_2}(
\nu^{\mu}_{k_2},\mbox{\boldmath $\rho$}^{\mu}_2)\right ]_L
\right ]_{JM}\right \},
\label{trialfnterm}
\end{equation}
where $\nu^{\mu}_{k_i}$ is the $k$th size parameter of the $i$th relative 
motion in the cluster arrangement $\mu$, ${\cal A}$ is the intercluster 
antisymmetrizer normalized such that the normalization kernel approaches 
the unit operator in the limit of infinite cluster separation,
$\Phi _{SM_S}$ is a product of the intrinsic wave
functions of the two $\alpha$-particles and the nucleon's spin-isospin 
function and $K$ stands for
the set of the indices $\{k_1,k_2\}$ of the size parameters.
By using an integral transformation \cite{VS},
the antisymmetrized product in Eq. (3) can be rewritten as a linear 
combination of
Slater determinants of Gaussian wave-packet single-particle functions. 
The matrix
elements between Slater determinants of these nonorthogonal single-particle
states are easily evaluated and can be expressed in a closed 
analytical form. 
\par\indent
The variational trial function is a combination of different arrangements and
intercluster angular momenta:
\begin{equation}
\Psi_{JM}=\sum_{(\ell_1\ell_2)L}\Bigl\{
\Psi^{\mu_1}_{[S,(\ell_1\ell_2)L]JM}+\Psi^{\mu_2}_{[S,(\ell_1\ell_2)L]JM}\Bigr\}.
\label{trialfn}
\end{equation}
It is noted that our wave function is fully antisymmetrized, free from 
the spurious center-of-mass motion (actually the total center-of-mass motion 
is eliminated) and has a good total angular momentum and parity. Our 
calculation is the so-called ``variation after projection'' type.
\par\indent
The partial waves in a given cluster arrangement form a complete set of
states and the different Jacobi coordinate systems are, therefore, 
equivalent in principle. One might thus think that we only need to choose a 
particular arrangement, $\mu_1$ or $\mu_2$, and to decompose the wave 
function into a complete 
set of partial waves in this arrangement, and that the inclusion of both 
the arrangements implied in Eq. (4) would be redundant. Our experience 
\cite{VST} 
shows, however, that the convergence of energy in a fixed arrangement is 
rather slow. The reason is that the components 
$\Psi^{\mu_1}_{[S,(\ell_1\ell_2)L]JM}$ and
$\Psi^{\mu_2}_{[S,(\ell_1'\ell_2')L']JM}$
in the arrangements $\mu_1$ and $\mu_2$ are rather 
different, especially, at large distances and that 
any component $\Psi^{\mu_1}_{[S,(\ell_1\ell_2)L]JM}$ 
can only be represented by an infinite sum in terms of the arrangement 
$\mu_2$.  Moreover, the inclusion of high partial waves in the calculation 
is quite expensive. Our favorite choice is, therefore, 
to (1) decompose the wave function into 
partial waves in a given arrangement, (2) truncate the
higher partial waves, and (3) complete the wave function by the inclusion
of low partial waves of different arrangements.
\par\indent
The arrangements and the angular momenta combined
with the size parameters in the expansion make the dimension of
the basis large. These basis functions are, however,
nonorthogonal to each other and not all of them are equally important.
In a previous paper \cite{VLS} we tested different methods to select
the parameters $\nu_{k_i}^\mu$ that span most adequately the state
space, while the dimension of the basis is kept feasible.
The most efficient procedure found is the stochastic selection 
\cite{VST,VS}: We 
generate size parameter sets by a random choice from a region which
is physically important. The parameter sets
that satisfy an admittance condition are 
chosen to generate
basis states. 
The calculation was repeated several times to check the convergence. 
The dimension for the $^9$Be ground state is around 90.

\subsection{The scaling method for resonances}
\par\indent
Except for the ground state of $^9$Be, all the states of $^9$Be and $^9$B 
are above the three-body threshold. The $\frac{1}{2}^+$, 1.68 MeV state of $^9$Be 
lies just 111 keV above the 
threshold, but has a width of 217$\pm$10 keV. The $\frac{5}{2}^-$, 2.43 MeV 
state has a narrow width. The widths of other states of isospin 
$\frac{1}{2}$ range from several hundreds keV to about 1 MeV. The states 
of $^9$B 
have generally wider widths than the corresponding states of $^9$Be. 
\par\indent
Resonances are associated with complex eigenvalues of the time-independent
Schr\"odinger equation. It is not trivial to calculate the energy 
and the width of a resonance state  for a complex system.  Several 
methods have been developed to obtain
these complex eigenvalues using square integrable functions. 
The most well-known methods are the complex scaling \cite{abc} and the 
stabilization \cite{stabili} methods. 
\par\indent
The complex scaling method uses the unitary transformation which 
dilates the internal coordinates of the system 
according to $x \rightarrow xe^{i\theta}$, making the resonant wave 
functions square integrable. The eigenvalues that are associated 
with metastable resonance states
appear as such complex eigenvalues that are independent of the 
scaling angle $\theta$,
when it is larger than a critical angle, and the eigenvalues that 
are associated with nonresonant continuum states appear as 
complex eigenvalues which are dependent on the scaling angle \cite{abc}. 
One can expand the eigenfunctions 
of the complex scaled Hamiltonian in terms of square integrable 
basis functions  as  we did for bound states.
The variation of the energy functional with respect to the trial 
function, however,
yields a stationary rather than a minimum principle. Therefore, the 
stochastic basis selection procedure cannot  be applied here, but 
instead, we will work on a basis with fixed nonlinear parameters. 
\par\indent
The stabilization method \cite{stabili} utilizes the discrete states 
calculated in a box of large size. The stabilization method can  
be combined with the stochastic variational method. In this case we 
select the basis parameters from a confined interval.
\par\indent
These methods have been widely applied for two- and three-body resonances 
in atomic physics. Recently, nuclear physicists have also began 
to use the complex scaling method as a useful tool to locate two- 
\cite{kruppa} and three-body \cite{csoto} resonances of nuclear systems.
\par\indent
Due to the complexity of the problem both methods require extreme 
numerical accuracy. To be on the safe side, we used these methods only 
when they are certainly able to give reliable results. 
That is, we used the stabilization method
for narrow resonances, and calculated only the resonance energy 
because the calculation
of the width would require an excessively large computational burden. 
For these quasibound states the stabilized wave functions can directly 
be used to calculate the matrix element of a physical operator because they 
are real. 
To locate wider resonances we used the complex scaling method. In this 
case we calculated both the width and the position.
\par\indent
We have found that the energy of the narrow $\frac{5}{2}^-$ state 
can well be 
obtained by diagonalizing the Hamiltonian in a sufficiently large 
basis of Eq. (3). The 
resonance energy remains rather stable against the change of the basis set 
within a reasonable range. The wave function obtained in this way is used to 
calculate the electromagnetic transition rates. It is very difficult to do 
better than this because enclosing the wave function in a box as required 
by the stabilization method is not trivial for the three-body system.  
\par\indent
To apply the complex scaling method to the present case, we define 
the transformation $U(\theta)$ which acts on the function of 
the intercluster Jacobi coordinates, \mbox{\boldmath $\rho$}$_1^\mu$ and 
\mbox{\boldmath $\rho$}$_2^\mu$, 
\begin{equation}
U(\theta)f(\mbox{\boldmath $\rho$}_1^\mu,\mbox{\boldmath $\rho$}_2^\mu ) 
= {\rm e}^{{3i\theta}/2}
f(\mbox{\boldmath $\rho$}_1^\mu \, {\rm e}^{i\theta}, 
\mbox{\boldmath $\rho$}_1^\mu \, {\rm e}^{i\theta} ).
\end{equation}
The eigenvalue problem of the transformed Hamiltonian $H_{\theta}=
U(\theta)HU(\theta)^{-1}$ is solved for each $\theta$ value. A resonance 
state corresponds to an square integrable solution of the transformed 
Hamiltonian and may be described as in Eq. (4).  
When the basis function of Eq. (3) is employed, the 
operation $U(\theta)^{-1}$ on the relative motion function is equivalent to 
multiplying the size parameters, $\nu^{\mu}_{k_1}$ and $\nu^{\mu}_{k_2}$, 
by ${\rm e}^{-2i\theta}$. The energy $E_R$ and the width $\Gamma_R$ of a 
resonance are obtained as the real and imaginary parts of a complex 
eigenvalue, $E_{\theta}=E_R-\frac{1}{2}i\Gamma_R$, of $H_{\theta}$,   
which remains unchanged for arbitrary values of $\theta$ within 
an appropriate range.

\section{RESULTS}
\subsection{Input parameters}
The internal state of  the $\alpha$-particle was approximated 
by 0s harmonic-oscillator Slater determinant
wave function of a size parameter $\nu$=$m\omega /2\hbar$. 
The value of $\nu$ was chosen to be 0.26 fm$^{-2}$ to reproduce the  
experimental charge radius of the $\alpha$-particle. 
The results are insensitive to the choice of the size parameter
within a reasonable limit.
\par\indent
  We used  Minnesota nucleon-nucleon interaction \cite{minnesota},
which is a sum of central and spin-orbit potentials of Gaussian form. The 
Coulomb potential was included. 
The strength of the spin-orbit force was taken from the set IV of 
Reichstein and Tang, which gives a good fit to $N+\alpha$ phase shifts. 
The central part of the Minnesota potential contains an exchange-mixture 
parameter $u$. The potential with $u$=1 corresponds to a Serber type 
mixture.  
Decreasing the value of $u$ from unity implies increasing repulsion in odd 
partial waves, while keeping the strength of even partial waves unchanged. 
It 
was set to $u=0.94$ in order to reproduce the ground state energy of $^9$Be. 
The value of $u=0.94$ is very close to the value (0.95) which is needed to 
well describe the $\alpha+\alpha$ scattering in the resonating group method 
\cite{minnesota}. 
Thus our choice should give a realistic interaction between the 
$\alpha$-particles. The value of $u=0.94$ is, however, slightly smaller 
than the value of 0.97 recommended for the description of $N+\alpha$ 
scattering.  By fixing the $u$ and $\nu$ parameters
as described above, the model contains no free parameter. No change of the 
potential parameters was made between $^9$Be and $^9$B.

\subsection{Cluster arrangements and angular momentum channels}
In our model the total spin is uniquely given by $S$=$\frac{1}{2}$ 
so that the total 
orbital angular momentum can take either $L=J-\frac{1}{2}$ or 
$L=J+\frac{1}{2}$. Let us 
show that both values of $L$ are needed by taking an example of 
the magnetic moment of 
$^9$Be. Quite probably (and this will be confirmed later) 
the orbital motion of the protons gives  
a moderate contribution to the magnetic moment of $^9$Be and 
only the spin part needs to be considered to get a reasonable estimate 
of the magnetic moment. The magnetic moment is then approximated by 
($J$=$\frac{3}{2}$, $L=1$ and $L=2$) 
\begin{eqnarray}
\mu&=&\langle \Psi_{JJ}|\mu_z|\Psi_{JJ} \rangle 
=g_s(n)\sum_L c_L^2 \left(\sum_{M_SM_L} \langle SM_SLM_L|JJ \rangle^2
M_S\right) \nonumber\\
&=&g_s(n)\sum_L c_L^2\left(\frac{[J(J+1)+S(S+1)-L(L+1)]J}{2J(J+1)}\right),
\end{eqnarray}
where $g_s(n)=-3.826$ is the spin $g$-factor of the neutron in units of 
nuclear magneton and $c_L$ is the amplitude 
of the total orbital angular momentum $L$ in the ground state wave function. 
If the ground state is purely of $L$=1, then the magnetic moment becomes 
$-1.913$ $\mu_N$, which is in disagreement with the observed value of 
$\mu_{\rm exp}=-$1.1778 $\mu_N$. An $L=$2 component of about 20 \% 
admixture is needed to reproduce the observed value. We will see later 
that the potential chosen gives just the needed admixture. It is 
instructive to note that the magnetic moment for pure $L$=1 case is equal 
to the Schmidt value of the single $p_{3/2}$ neutron. 
\par\indent
Table I lists a set of arrangements and angular momenta used in the 
present calculation.  
We did several pilot calculations to know the relative importance of the 
arrangements and the angular momentum channels. When all the nine sets 
of Table I are used for the $\frac{3}{2}^-$ ground state, the energy 
from the 
$\alpha+\alpha+n$ threshold is obtained as $-1.431$ MeV and the root mean 
square (rms) radius of point nucleon is 
2.50 fm. Let us call this a full calculation. When we exclude three sets  
belonging to the arrangement $\mu_2=(\alpha N)\alpha$ with $\ell_1=0$ or 2, 
both energy and 
radius hardly change from the result of the full calculation; the overlap 
of the approximate wave function with the full wave function is 0.9995. 
This result is physically acceptable because the $p$ wave is of prime  
importance for the interaction between the neutron and the 
$\alpha$-particle. If we further exclude three sets belonging to the $\mu_2$ 
arrangement with $\ell_1=1$, then the energy increases to $-0.32$ MeV and 
the 
radius increases to 2.57 fm. This suggests that the arrangement $\mu_1
=(\alpha\alpha)N$ 
($^8$Be+$n$-type configuration) alone is imperfect to describe 
the ground state even though the $s$ and $d$ waves are taken into account 
for the motion of the two $\alpha$-particles. This consideration leads us 
to the remark that the calculations of Refs. \cite{voron,desc} using 
only the 
$^8$Be+$N$ channel should be accepted with some reservations. On the other 
hand, if we exclude  
three sets belonging to the $\mu_1$ arrangement, then the 
result is very 
close to the full calculation; the energy loss is merely 34 keV and the 
overlap of the wave functions is 0.9991. We can thus conclude that the 
$^5$He+$\alpha$-type configuration with $\ell_1$=1 constitutes a very good 
approximation to the ground state wave function. As is seen from Table I, 
the angular momentum in the $\mu_2$ arrangement is restricted to 
$\ell_1=1$ for other states. 
\par\indent
For resonance states, particularly for high spin resonances 
the inclusion of high partial waves becomes important to obtain stable  
resonance parameters in the complex scaling method. 
The complex eigenvalue of the rotated Hamiltonian $H_{\theta}$ is 
obtained by using the basis function of Eq. (3). The size parameters of 
the basis function 
are not selected randomly but are chosen as $\nu^{\mu}_{k}=\nu_{0}p^{k-1}$ 
$(k=1,...,K)$. The values of $\nu_{0}$ and $K$ are varied for each resonance 
to get stable values for its energy and width. The adopted value of $K$ is 
about 
10 in the present calculation. The basis dimension used to diagonalize the 
rotated Hamiltonian is $K^2$ times the number of the sets listed in Table I. 
Figure 2 displays an example of the complex scaled spectra of $^9$Be 
for $J^{\pi}=\frac{3}{2}^{-}$ and $\frac{7}{2}^{-}$. One can see, besides 
the 
discretized points corresponding to the three-body continuum, those points 
which lie on straight lines starting from the positions of the resonances 
of the subsystems. 

\subsection{Energy spectrum and electromagnetic properties}
The calculated spectra of $^9$Be and $^9$B are compared with experiment  
in Fig. 3. The theoretical level sequence in $^9$Be has a good 
correspondence with the observed 
spectrum. The second $\frac{3}{2}^-$ resonance is obtained at 4.3 MeV 
excitation energy. The other calculations 
\cite{okabe,suppl,desc} also predict the $\frac{3}{2}^{-}_{2}$ state. 
Although no such state is cited in Ref. \cite{ajzen}, the calculated 
resonance may correspond to the state at 5.59 
MeV mentioned in Ref. \cite{dixit}.  We get two broad overlapping resonances 
with $\frac{7}{2}^-$ and $\frac{9}{2}^+$ at about 6.5 MeV. This   
agrees with the conclusion of the recent experiments \cite{glickman,dixit}. 
We could not find a resonance with $\frac{1}{2}^-$ around 8 MeV 
excitation energy in accordance with Refs. \cite{glickman,dixit}, 
although such a state is parenthetically quoted 
in Ref. \cite{ajzen}. Instead of this a $\frac{5}{2}^-$ resonance 
is obtained at 7.9 MeV, which agrees with the result of Refs. 
\cite{okabe,suppl}.  The spectrum 
of $^9$B is less known experimentally compared to that of $^9$Be. 
The calculated spectrum is similar to the one of $^9$Be.  
We can predict the energy and the width of several resonances in 
$^9$B with the same accuracy as the case of $^9$Be.  For example, our 
calculation predicts a missing $\frac{1}{2}^-$ state at 2.43 MeV, which is 
in agreement with the result of a recent $^9$Be($p,n$) 
reaction \cite{pugh} that located the $\frac{1}{2}^-$ state at 2.83 MeV. 
Although no definitive spin assignment is made to the state at 2.788 MeV 
excitation energy \cite{ajzen}, our calculation supports a 
$\frac{5}{2}^+$ assignment rather than $\frac{3}{2}^+$.  
\par\indent
Table II lists the energies and the widths of the unbound states 
calculated by the complex scaling method. The energies of the 
$\frac{5}{2}^-$ states of both $^9$Be and $^9$B are in good agreement with 
experiment. Their widths, though extremely narrow, are reasonably 
reproduced; the calculated width 
of $^9$Be is about 2 times larger than the observed value, while the width 
of $^9$B is about a half of the experiment. The calculation reproduces 
the widths of other states within a factor of 2. Our result is in better 
agreement with experiment than the calculation of Ref. \cite{desc}. 
\par\indent
There has been considerable effort to determine the location of the 
$\frac{1}{2}^+$ state from the point of view of a Thomas-Ehrman shift 
\cite{shift}. We applied the complex scaling method to find a resonance 
with  $J^{\pi}=\frac{1}{2}^+$ by including the arrangements and the angular 
momentum channels listed in Table I. The present calculation could 
not identify such a stable complex 
eigenvalue that can be interpreted as a resonance. 
To estimate the $E$1 transition strength, we increase the 
value of $u$ to make the $\frac{1}{2}^+$ state particle-bound.
\par\indent
The electromagnetic moments and the rms radii of proton, neutron, and 
nucleon, assuming pointlike nucleons, are included in Table III. 
Bare operators are used in the calculation. 
The charge radius of $^9$Be with the effect of the proton's finite size 
becomes 2.54 fm and fits the experimental value of $2.519\pm0.012$ fm 
\cite{ajzen}. 
The rms radius of neutron is larger than that of proton by 0.2 fm. Both 
the magnetic and the  
quadrupole moments of $^9$Be are reproduced very well. As was stated in 
subsection IIIB, the contribution of the proton's orbital motion to the 
magnetic moment is rather small (0.28 $\mu_N$) and the contribution of the 
spin part, $-1.45 \,\mu_N$, corresponds to 15.1 \% admixture of the $L=$2 
component. 
The $M1$ and $E2$ transition probabilities of the $\frac{5}{2}^-$ state 
to the ground state are also well reproduced. The strong $E1$ transition 
of the $\frac{1}{2}^+$ state is in reasonable agreement with experiment. The 
$E1$ transition strength depends on the description of the tail part of the 
wave function. With $u=1.0$ the energy of the $\frac{1}{2}^+$ state 
changes to 
593 keV below the threshold and the $B(E1)$ value becomes 0.24 W.u. in 
good agreement with 
experiment. With $u=0.98$ the energy goes up to 206 keV below the threshold 
and the exterior part of the wave function that does not contribute to the 
transition grows, thereby reducing the $B(E1)$ value to 0.18 W.u. 
To our best knowledge, this is the first theoretical calculation which 
has been able to reproduce the $E1$ transition probability in a consistent 
way.   
Reference \cite{barker} argues that the experimental $E1$ strength is 
enhanced to 0.38$\pm$0.07 W.u. if the unbound nature of the state is 
taken into account. 
\par\indent
Table III includes the results of other models. 
The $\mu$ and $Q$ moments of the shell model 
were determined by using an effective interaction which 
was chosen to reproduce both energies and static moments of 
$0p$-shell nuclei \cite{hees84}. These values are rather close to those of 
Cohen-Kurath (8-16) POT calculation \cite{cohen,glickman}. A shell-model 
calculation 
of (0+1)$\hbar \omega$ model space \cite{hees84} cannot account for the 
enhancement of the $B(E2)$ transition; 
with the effective charge of 0.35$e$ it gives about one third of the 
experimental value. The $E1$ transition probability of the lowest 
$\frac{1}{2}^{+}$ state to the ground state was predicted to be only 
0.03 W.u. \cite{hees84}. Another shell-model calculation in a similar basis 
\cite{glickman} reproduces reasonably the $B(E2)$ value by using a large 
effective charge for neutron, but again gives a very small $B(E1)$ value. 
Although the calculation of Ref.  
\cite{desc} using only $^8$Be+$n$ channel gives a reasonable agreement 
with experiment, we have already pointed out that the $^5$He+$\alpha$ type 
configuration leads to further improvement. A similar remark applies to the 
calculation of Ref. \cite{voron}, which indicates that the charge radius and 
the quadrupole moment are considerably smaller than experiment. 
\par\indent
Further test of the wave function of the $^9$Be ground state is performed  
by the electron scattering data \cite{deforest}. The longitudinal 
electron scattering form factor is 
calculated in a first-order plane wave Born approximation through 
\begin{equation}
|F_L(q)|^2=\frac{4\pi}{Z^2(2J_i+1)}\sum_{\ell}|\langle\Psi_{J_f}\|\hat{M}
_{\ell}^{\rm coul}(q)\|\Psi_{J_i}\rangle|^2,
\end{equation}
where $Ze$ is the charge of the nucleus and the reduced matrix element 
of the operator $T_{\kappa}^k(q)$ is defined by 
\begin{equation}
  \langle JM|T_{\kappa}^k|J'M' \rangle=\frac{(-1)^{2k}}{\sqrt{2J+1}}
  \langle J'M'k \kappa |JM \rangle \langle J\|T^k\|J'\rangle .
\end{equation}
The charge density multipole operator $\hat{M}_{\ell m}^{\rm coul}(q)$ 
which occurs 
in the form factor is given as a function of momentum transfer $q$ from the 
charge density operator 
\begin{equation}
 \hat{M}_{\ell m}^{\rm coul}(q)=\int j_{\ell}(qr)Y_{\ell m}
({\hat{\bf r}})
 \sum_{i=1}^A
 \frac{1-\tau_{3_i}}{2} \delta({\bf r}_i - 
{\bf R}_{\rm cm} - {\bf r}) 
d {\bf r} ,
\end{equation}
where {\bf r}$_i$ is the nucleon coordinate and 
{\bf R}$_{\rm cm}$ is the total 
center-of-mass coordinate. Note that our wave function contains 
no center-of-mass motion. Figure 4 compares the calculated charge 
form factor with the experiment \cite{meyer,bern,slight}. 
The correction for the finite proton size is taken into account by 
multiplying the form factor with the proton's form factor used in Ref. 
\cite{protonff}. Both monopole (C0) and quadrupole (C2) 
terms contribute to the charge 
form factor. No effort has so far been made to separate those contributions 
experimentally. Polarized electrons and targets will be needed to do 
such experiments. The agreement between theory and experiment 
is good. This is perhaps not very surprising because the present model 
reproduces both charge radius and quadrupole moment accurately. It is clear 
that the quadrupole deformation of the charge density 
is important at higher $q^2$ values. The deformation of the proton and 
neutron density distributions will be discussed in the next subsection. 
\par\indent
The transverse electron scattering form factor gives information on the 
nuclear current density. It is calculated from the expression 
\begin{equation}
|F_T(q)|^2=\frac{4\pi}{Z^2(2J_i+1)}\sum_{\ell}\Bigl\{|\langle\Psi_{J_f}\|
\hat{T}
_{\ell}^{\rm el}(q)\|\Psi_{J_i}\rangle|^2 + |\langle\Psi_{J_f}\|
\hat{T}
_{\ell}^{\rm mag}(q)\|\Psi_{J_i}\rangle|^2 \Bigr\}.
\end{equation}
The symmetry consideration on parity and time reversal shows that the 
elastic form factor receives no contribution of 
the transverse electric multipoles of the current density 
$\hat{\mbox{\boldmath $j$}}$({\bf r}). The transverse 
magnetic multipoles are defined by
\begin{equation}
 \hat{T}_{\ell m}^{\rm mag}(q)=\int j_{\ell}(qr)\mbox{\boldmath $Y$}
_{\ell \ell 1}^m(\hat
{\bf r})\cdot \hat{\mbox{\boldmath $j$}}({\bf r})
d {\bf r}.
\end{equation}
Here the vector spherical harmonics are defined with unit vector 
{\bf e} as 
\begin{equation}
 \mbox{\boldmath $Y$}_{\ell' \ell 1}^m(\hat{\bf r})=
 [Y_{\ell}(\hat{\bf r}){\bf e}]^{\ell'}_m
\end{equation}
and the current density consists of the convection and magnetization 
currents:
\begin{eqnarray}
\hat{\mbox{\boldmath $j$}}({\bf r})&=&\frac{1}{2mc}\sum_{i=1}^A
 \frac{1-\tau_{3_i}}{2} \Biggl\{ {\bf p}_i\delta(
{\bf r}_i - {\bf R}_{\rm cm} - 
{\bf r})+ \delta({\bf r}_i - 
{\bf R}_{\rm cm} - {\bf r}){\bf p}_i
 \Biggr\} \nonumber\\
&+&\nabla\times\Biggl(\frac{\hbar}{2mc}\sum_{i=1}^A\mu_i\delta({\bf r}_i - 
{\bf R}_{\rm cm} - {\bf r})\mbox{\boldmath 
$\sigma$}_i\Biggr).
\end{eqnarray}
Here {\bf p}$_i$ is the momentum of the nucleon in the center-of-mass 
system and $\mu_i$ is the magnetic moment of the nucleon in units of nuclear 
magneton. Figure 5 compares the calculated transverse form factor with 
the data of Refs. \cite{rand,lapikas}. Both $M1$ and $M3$ 
contributions 
are important to get a satisfactory reproduction of experiment. Shell-model 
calculations \cite{glickman} needed a quenching factor of about 0.7 for 
the transverse form factors, while no quenching is needed in our model.  
We can conclude that the ground-state wave function of 
the present model reproduces consistently all the electromagnetic 
properties of $^9$Be.

\subsection{Density distributions and spectroscopic amplitudes}
The proton and the neutron  density distributions, defined by
\begin{equation}
\rho({\bf r})=\Bigl\langle \Psi_{JJ} \Bigm\vert \sum_{i=1}^{A} 
\delta ({\bf r}_i-{\bf R}_{\rm cm}-{\bf r})
P_i \Bigm\vert  \Psi_{JJ} \Bigr\rangle = \rho_0(r)+ \sum_{\ell \ne 0} 
\rho_{\ell}(r) Y_{\ell 0}({\hat {\bf r}}),
\end{equation}
(where $P_i$ projects out the protons or neutrons) are also
determined. For the ground state of $^9$Be we have monopole 
and quadrupole($\ell=2$) densities. The density distributions,
$\rho_0(r)$ and $\rho_2(r)$, are shown in Figs. 6(a) and 6(b). They are 
related 
to the rms radius and the quadrupole moment as below:
\begin{eqnarray}
\langle r^2 \rangle&=&\frac{4\pi}{Z}\int_0^{\infty}
\rho_0(r)r^4 dr,\\
Q&=&(\frac{16\pi}{5})^{1/2}\int_0^{\infty}\rho_2(r)r^4 dr.
\end{eqnarray} 
An analogous relation can be defined for the neutron case. The quadrupole 
moment becomes 5.13 fm$^2$ 
for the proton and 3.86 fm$^2$ for the neutron. The fact that the neutron 
quadrupole moment is smaller than the proton quadrupole moment is 
understood by noting that the single neutron cluster moves between 
the two 
$\alpha$-particles for the most of time and thus makes the neutron density 
less deformed. 
\par\indent
The 2.43 MeV, $\frac{5}{2}^-$ and 6.38 MeV, $\frac{7}{2}^-$ states of 
$^9$Be together with the ground 
state approximately follow a $J(J+1)$ rule and can be considered to 
form a rotational band with $K$=$\frac{3}{2}$ \cite{glickman,dixit}. 
>From the experimental quadrupole moment of the 
ground state, the intrinsic quadrupole moment $Q_0$ of the band is estimated 
as 26.5 fm$^2$ by using the relation $Q=\frac{J(2J-1)}{(J+1)(2J+3)}Q_0=
\frac{1}{5}Q_0$ 
\cite{bohr}.  The $E2$ transition probability within the band is related,  
in the collective model, to the $Q_0$ value by 
\begin{equation}
B(E2;KJ_1 \to KJ_2)=\frac{5}{16\pi}e^2Q_0^2\langle J_1K20|J_2K \rangle^2,
\end{equation}
which predicts 23.9 $e^2$fm$^4$ and 9.98 $e^2$fm$^4$ for the 
$\frac{5}{2}^-\to \frac{3}{2}^-$ 
and $\frac{7}{2}^-\to \frac{3}{2}^-$ transitions, respectively. The 
corresponding experimental values are 27.1$\pm$2.0 $e^2$fm$^4$ 
and 7.0$\pm$3.0 $e^2$fm$^4$ \cite{ajzen}. 
Since the collective model prediction agrees reasonably well with 
experiment, it may be possible to extract the intrinsic deformation 
parameter $\beta_0$ by using the relation $\beta_0=\sqrt{\frac{\pi}{5}}
\frac{Q_0}{Z\langle r^2 \rangle}$. Our theory gives $\beta_0$=0.89, which is 
close to the empirical deformation parameters of neighbouring nuclei, e.g., 
  $\beta_0\sim$1.13 for $^{10}$Be and $\beta_0\sim$0.82 for $^{10}$C,  
while the corresponding $Q_0$ values are 22.9 fm$^2$ and 25.0 
fm$^2$, respectively \cite{raman}. The deformation parameter $\beta$ 
associated with 
the density of Eq. (14) is estimated by assuming that it can be 
approximated by 
$\rho_{s}(r/(1-\frac{1}{4\pi}\beta^2+\beta Y_{20}({\hat {\bf r}}))$ 
from a spherical shape $\rho_{s}(r)$. The extracted value of 
$\beta$ is close to 1/5 of the $\beta_0$ value as expected by the 
collective model. 
\par\indent
The monopole densities of the protons and the neutrons may be used to 
calculate the total reaction cross section at high energies. It is given, 
in the Glauber theory \cite{glauber}, as 
\begin{equation}
\sigma_{R}=\int [1-\exp\{-2{\rm Im}{\chi}({\bf b})\}] d {\bf b},
\end{equation}
where {\bf b} is the impact parameter and the phase shift function, 
$\chi$, is related to the densities of the target and the projectile through 
the thickness function, $T({\bf s})=\int \rho({\bf s},z)dz$, by 
\begin{equation}
i\chi({\bf b})=-\int\!\!\!\int T_P({\bf s})T_T({\bf t})
\Gamma({\bf b}+{\bf s}-{\bf t})d{\bf s}d{\bf t}.
\end{equation}
Here $\Gamma$ is the profile function for the $NN$ scattering. The 
monopole densities of the proton and the neutron were used to construct 
the density of $^9$Be. The $\sigma_{R}$ value of $^9$Be for a 
Carbon target at 800 MeV/nucleon is calculated to be 850 mb with the 
parametrization of the profile function used in Ref. \cite{ogawa}. The 
interaction 
cross section measured by Tanihata {\it et al}. \cite{tanihata} is 
not exactly the same as 
but approximately equal to the reaction cross section. Their value is 
806$\pm$9 mb, which is in a fair agreement with theory. The reaction cross 
section of 
$^9$Be on a Cu target was measured by Saint-Laurent {\it et al}. 
\cite{saint} at about 45 
MeV/nucleon. They extracted the reduced strong absorption radius, 
$r_0\sim1.13$ fm, 
for $^9$Be by fitting their measured cross sections to the formula by 
Kox {\it et al} \cite{kox}. 
This formula predicts $\sigma_{R}=825\pm20$ mb for the $^9$Be+$^{12}$C 
system at relativistic energies as listed in Table III. We again confirm 
that our density is reliable enough to reproduce the experiment. 
\par\indent
We showed in the previous subsection that the enhancement of the $E1$ 
transition of the first excited state in $^9$Be is reproduced well. 
To understand this we note that the $E1$ operator is recast to  
$\frac{NZ}{A}e\sqrt{\frac{3}{4\pi}}({\bf R}_Z - {\bf R}_N)$, 
where ${\bf R}_Z$ and ${\bf R}_N$ are  the center-of-mass 
coordinates of the protons and the neutrons, respectively. 
The enhancement of the transition 
should be therefore related to the excitation of the corresponding motion 
in the excited state. In the $\alpha+\alpha+n$ model the $E1$ excitation 
is caused by the valence neutron. Figure 7 compares the monopole density 
of the $\frac{1}{2}^+$ state obtained with $u=1.0$ with that of the 
ground state. The proton 
density becomes smaller near the center but reaches up to larger distances, 
indicating the increase in the mean distance between the two 
$\alpha$-particles. The neutron density shows a substantial decrease at 
1$\sim$2 fm and a significant increase beyond 3 fm. The proton and 
neutron rms radii increased from 2.39 to 2.94 fm and from 2.58 to 
5.59 fm, respectively. Though the increase of the proton size is moderate, 
that of the neutron size is dramatic. The picture emerging 
from this analysis is the following: The valence neutron in the ground 
state is mostly confined between the two $\alpha$-particles but, in the 
excited $\frac{1}{2}^+$ state, moves around them in a spatially wider 
region. 
It is easy to understand that the large $E1$ transition strength has 
naturally come out from the structure change of the underlying states. 
\par\indent   
Another interesting quantity that helps to reveal information on the wave 
function is the spectroscopic amplitude which, in the angular 
momentum projected basis, is defined by 
\begin{equation}
g^{\mu}_{(\ell_1\ell_2)L}(r,R)=\frac{1}{r^2R^2}\langle {\cal A}\left
\{\left [\Phi _S
\left [Y_{\ell_1}({\hat{\mbox{\boldmath $\rho$}}}^{\mu}_1)Y_{\ell_2}(
{\hat{\mbox{\boldmath $\rho$}}}^{\mu}_2)\right ]_L
\right ]_{JM} \delta(\rho^{\mu}_1-r) \delta(\rho^{\mu}_2-R)\right \}|
\Psi_{JM}\rangle.
\end{equation}
Figures 8(a), 8(b), and 8(c) display the spectroscopic amplitudes of the 
ground state of $^9$Be for some channels of the 
arrangement $\mu_1=(\alpha\alpha)n$, letting $r$ and $R$ represent 
the distances of the two $\alpha$-particles and of the neutron from their 
center-of-mass, respectively. Some remarkable features are that all the 
amplitudes have a peak at $r\sim3.2$ fm and $R\sim2.3$ fm and 
$R$-independent nodes at $r=1$ fm (for the $s$ wave between $\alpha$'s) and 
$r=2$ fm (for the $s$ and $d$ waves). These characteristics are understood 
by the microscopic $\alpha$-$\alpha$ cluster-model analysis for $^8$Be.   
The appearance of the nodes is understood in relation to the existence of 
the Pauli-forbidden states \cite{saito}. The norm of the amplitude, which 
is called the spectroscopic factor, 
becomes 1.03, 0.77, and 0.32 corresponding to three channels shown in 
Fig. 8. We note that the norm is different from the so-called amount 
of clustering. The amplitudes corresponding to the arrangement $\mu_2=
(\alpha n)\alpha$ are plotted in Fig. 9, where $r$ is now the distance 
between $n$ and $\alpha$ and $R$ the distance between their center-of-mass 
and $\alpha$. The nodes appear also in this case but their positions alter 
particularly at large $r$. This is due to the fact that $R$ is 
approximately equal to the $\alpha$-$\alpha$ distance at small $r$ but 
deviates largely from it with increasing $r$. The spectroscopic factor is 
0.84 and 0.61, respectively.  

\subsection{Beta decay of the $^9$Li ground state to $^9$Be}
Because the ground state and the $\frac{5}{2}^{-}$, 2.43 MeV state are 
described well by the present model, the $\beta$ decay of the $^9$Li ground 
state to these states is expected to further test the accuracy of their 
wave functions or an available wave function of $^9$Li. The experimental 
value of log$ft$ for the $\beta$ decay to the $^9$Be ground state is about 
5.31 \cite{ajzen,nyman}, indicating that the $\beta$-decay matrix element is 
fairly suppressed despite the allowed transition. The weak $\beta$ decay 
is ascribed to the fact that the spatial symmetry of the main component 
of $^9$Be is different from that of $^9$Li \cite{cohen}. The 
Gamow-Teller (GT) matrix element, 
\begin{equation}
M_{GT}(i \to f)=\Bigl\langle\Psi_{J_f}(^9{\rm Be})\Bigm\|\sum_{k=1}^{9}
t_{-}(k)
\mbox{\boldmath $\sigma$}(k)\Bigm\|\Psi_{J_i}(^9{\rm Li})\Bigr\rangle,
\end{equation}
to any state of $^9$Be, if it is described by the $\alpha+\alpha+n$ 
three-cluster model, always vanishes regardless 
of the wave function of $^9$Li. This is most easily understood by acting the 
Hermitian conjugate of the GT operator on the $^9$Be wave function and by 
noting that the spin-isospin part of the $\alpha$-particle wave function 
is fully occupied. 
\par\indent
The above discussion indicates that the simple three-cluster model for $^9$Be 
must be modified to explain the $\beta$ decay in spite of the successful 
results obtained in the previous subsections. The modification must be 
small enough not to destroy the agreement between experiment and the 
three-cluster model calculation. One possible way for the modification is 
to introduce the distortion of the $\alpha$-particle into $t+p$ and  
$h+n$ configurations. To explore the consequence of this modification, 
let us assume that the intrinsic wave function of the $\alpha$-particle 
can be expressed by 
\begin{equation}
\phi_{\alpha}=\sqrt{1-c^2}\phi_{\alpha}^{(0)} + c\phi_{\alpha}^{\rm (e)},
\end{equation}
where $\phi_{\alpha}^{(0)}$ represents the part which can be described by 
the 0s harmonic-oscillator Slater determinant, while 
$\phi_{\alpha}^{\rm (e)}$ the distorted part which is orthogonal to 
$\phi_{\alpha}^{(0)}$. The $^9$Be wave function in a more realistic 
three-cluster model can therefore be approximated by 
\begin{equation}
|\Psi_{J_f}(^9{\rm Be})\rangle = {\cal N}\left
\{(1-c^2)|\Psi_{J_f}^{(0)}(^9{\rm Be})\rangle 
+ 2c\sqrt{1-c^2}|\phi_{\alpha}^{(0)}\phi_{\alpha}^{\rm (e)}n\rangle + 
c^2 |\phi_{\alpha}^{\rm (e)}\phi_{\alpha}^{\rm (e)}n\rangle \right \}.
\end{equation}
Here the normalization constant, ${\cal N}$, is close to unity when $c$ is 
small and it is suppressed below. The first term, 
$|\Psi_{J_f}^{(0)}(^9{\rm Be})\rangle=|\phi_{\alpha}^{(0)}
\phi_{\alpha}^{(0)}n\rangle$, is nothing but 
the one described by the $\alpha+\alpha+n$ model and has no contribution to 
the $\beta$ decay. By neglecting the last term, the GT matrix element is 
given by 
\begin{eqnarray}
M_{GT}(i \to f)&=&2c\sqrt{1-c^2}\Bigl\langle\phi_{\alpha}^{(0)}
\phi_{\alpha}^{\rm (e)}n\Bigm\|\sum_{k=1}^{9}t_{-}(k)
\mbox{\boldmath $\sigma$}(k)\Bigm\|\Psi_{J_i}(^9{\rm Li})\Bigr\rangle 
\nonumber \\
&=& 2\langle\Psi_{J_f}^{(0)}(^9{\rm Be})|\Psi'_{J_f}(^9{\rm Be})\rangle
\Bigl\langle\Psi'_{J_f}(^9{\rm Be})\Bigm\|\sum_{k=1}^{9}t_{-}(k)
\mbox{\boldmath $\sigma$}(k)\Bigm\|\Psi_{J_i}(^9{\rm Li})\Bigr\rangle
\end{eqnarray}
with
\begin{equation}
|\Psi'_{J_f}(^9{\rm Be})\rangle = \sqrt{1-c^2}|\Psi_{J_f}^{(0)}(^9{\rm Be})
\rangle + c |\phi_{\alpha}^{(0)}\phi_{\alpha}^{\rm (e)}n\rangle .
\end{equation}
Equations (24) and (25) are our basic equations to calculate the $\beta$ 
decay matrix element when the distortion of the $\alpha$-particle is 
included. 
\par\indent
The wave function of eq. (25) is obtained by extending the three-cluster 
model to the four-cluster model which includes $\alpha+t+p+n$ and 
$\alpha+h+n+n$ partitions. 
In order to avoid excessive numerical calculations, the angular 
momentum channels and the cluster arrangements are rather limited. See 
Fig. 1(b) and Table IV.    
The spins of all the clusters were coupled to $S=\frac{1}{2}$. 
The isospins were not coupled to a definite value so that the wave function 
of the extended model may in general contain the total isospin of 
$T=\frac{1}{2}$ and $\frac{3}{2}$. The potential favors $T=\frac{1}{2}$ 
for the low-lying states of $^9$Be.      
\par\indent
The intrinsic wave function of the $t$- and $h$-cluster was described by 
0s harmonic-oscillator Slater determinant of the same size parameter $\nu$ 
as that of the $\alpha$-particle. 
The ground state wave function obtained in the four-cluster model using the 
Minnesota potential of $u$=0.94 has the overlap integral of 0.971  
with the one obtained in the three-cluster model. Therefore, this new 
wave function should yield substantially the same results as the 
previous one for the electromagnetic properties. This is just what 
we have expected to maintain in extending the model space. 
\par\indent
To calculate the $\beta$-decay probability we use the $^9$Li ground-state 
wave function which was obtained in a microscopic $\alpha+t+n+n$ 
four-cluster model \cite{VST}. This model for $^9$Li reproduced both  
magnetic and quadrupole moments of the ground state very well. To fit the 
energy of the $^9$Be ground state to its experimental value from the 
four-body threshold, 
we changed the $u$ parameter to 0.88 in the four-cluster 
model calculation. The overlap integral 
of the wave functions between the three-cluster and four-cluster 
models becomes 0.973 and the resulting log$ft$ value is 5.60. 
The log$ft$ value is still a little too large compared to the 
experimental value of 5.31, but this calculation 
strongly indicates that we are on the right track. A further refined 
four-cluster model calculation for both $^9$Be and $^9$Li will 
reduce the disagreement between experiment and theory because such 
a calculation is expected to enhance the GT matrix element. 
The shell-model calculation \cite{mikolas} gives the log$ft$ value 
in the range of 4.86 to 5.64, depending on the interaction used.         
It is interesting to analyze in the way presented above the $\beta$ 
decay of $^9$C to the low-lying 
states of $^9$B and an asymmetry in the $\beta$-decay matrix 
elements of $A=9$ nuclei \cite{mikolas,nyman}. 

\section{Summary}
\par\indent
The microscopic multicluster model was applied to the study of the mirror 
nuclei $^9$Be and $^9$B. They were described in a three-cluster model 
comprising two $\alpha$-particles and 
a single nucleon. The two-nucleon interaction consists of the central 
and spin-orbit potentials together with the Coulomb potential. The same 
two-nucleon potential was employed for both $^9$Be and $^9$B. The 
ground state of $^9$Be, an only 
particle-bound state in this study, was obtained with the stochastic 
variational method, while the other particle-unbound states were studied 
by the complex scaling and the stabilization  
methods. The three-body dynamics of the clusters was taken into account by 
including both of the arrangements, $(\alpha\alpha)N$ and 
$(\alpha N)\alpha$, and by using relevant partial 
waves between the relative motion of the clusters. The calculated spectrum 
of $^9$Be below an excitation energy of 8 MeV was in fair agreement with 
experiment. We obtained two broad overlapping resonances with 
$J^{\pi}=\frac{7}{2}^-$ and $\frac{9}{2}^+$ around 6.5 MeV, in 
agreement with the conclusion of the recent experiments. Two states,  
$\frac{3}{2}_{2}^-$ and $\frac{5}{2}_{2}^-$, were predicted 
at about 4.5 MeV and 8MeV in excitation energy, respectively. The 
spectrum of $^9$B was found to be similar to that of $^9$Be. The spin and 
parity of several states of $^9$B were predicted. The first excited 
$\frac{1}{2}^+$ state was not localized 
in the present study and thus no definite argument was possible on a  
Thomas-Ehrman shift in this case. 
\par\indent
The theory reproduced very well the electromagnetic properties of the $^9$Be 
ground state such as the charge radius, the magnetic moment, the quadrupole 
moment, and the elastic electron scattering form factors. The calculated 
ground state density was consistent with the total reaction cross section 
data. The intrinsic deformation parameter of the density was found to be 
0.89. The $\frac{1}{2}^+\to \frac{3}{2}^-$ $E1$ transition and the 
$\frac{5}{2}^-\to \frac{3}{2}^-$ $E2/M1$ transitions were studied by 
treating the excited states as bound. The calculated 
transition rates were in good agreement with experiment. 
\par\indent
The fact that the present calculation reproduced all the data well 
strongly supports that the three-cluster model is quite appropriate for 
describing the structure of $^9$Be and $^9$B, provided a full account 
of the dynamics is taken into account in the calculation. We were also 
able to understand the $\beta$ decay of $^9$Li to $^9$Be by admixing 
the small components that are induced by the distortion of the 
$\alpha$-particle into $t+p$ and $h+n$ configurations. A unique advantage 
of the microscopic multicluster model was exemplified by being able to 
accommodate such distortion in the model consistently. The study on heavier 
Be isotopes is in progress in the framework of the microscopic multicluster 
model including two $\alpha$-particles and several neutrons.       

\bigskip 
This work was supported  by a
Grant-in Aid for Scientific Research (No. 05243102 and No. 0664038) of the
Ministry of Education, Science and Culture (Japan) and by OTKA Grant No. 
T17298 (Hungary). Most of the calculations were done with the use of 
RIKEN's VPP500 computer.

\begin{table}
\caption{A set of arrangements and angular momenta included in the 
three-cluster model calculation for $^9$Be $(N=n)$ and $^9$B $(N=p)$. 
See Fig. 1(a) for the angular momenta $\ell_1$ and $\ell_2$.}
\end{table}

\vspace {0.5mm}

\begin{tabular}{p{1.5cm}p{3cm}p{1.7cm}p{1.7cm}p{1.7cm}p{1.7cm}p{1.7cm}
p{1.7cm}}
\hline\hline

$J^{\pi}$ & arrangement & 
 \multicolumn{5}{c}{angular momentum $(\ell_1,\ell_2)L$}   \\ \hline

$1/2^-$ 
  & $(\alpha \alpha)N$   & (0,1)1 & (2,1)1 & (2,3)1 &    &    &   \\
  & $(\alpha N)\alpha$   & (1,0)1 & (1,2)1 &        &    &    &  \\ \hline   

$1/2^+$ 
  & $(\alpha \alpha)N$   & (0,0)0 & (2,2)0 & (2,2)1 &    &    &  \\
  & $(\alpha N)\alpha$   & (1,1)0 & (1,1)1 &        &    &    &  \\ \hline

$3/2^-$   
   & $(\alpha \alpha)N$  & (0,1)1 & (2,1)1 & (2,1)2 &    &    &   \\
   & $(\alpha N)\alpha$  & (0,1)1 & (1,0)1 & (2,1)1 & (1,2)1 & (2,1)2 
                         & (1,2)2    \\ \hline

$3/2^+$   
   & $(\alpha \alpha)N$  & (2,2)1 & (0,2)2 & (2,0)2 & (2,2)2  &  (2,4)2 
                         & (4,2)2    \\
   & $(\alpha N)\alpha$  & (1,1)1 & (1,1)2 & (1,3)2 &    &    &   \\ \hline

$5/2^-$   
   & $(\alpha \alpha)N$  & (2,1)2 & (2,3)2 & (0,3)3 & (2,1)3  & (2,3)3 & \\
   & $(\alpha N)\alpha$  & (1,2)2 & (1,2)3 &        &    &    &   \\  \hline

$5/2^+$   
   & $(\alpha \alpha)N$  & (0,2)2 & (2,0)2 & (2,2)2 & (2,2)3  &  &  \\
   & $(\alpha N)\alpha$  & (1,1)2 & (1,3)2 & (1,3)3 &    &    &   \\  \hline

$7/2^-$   
   & $(\alpha \alpha)N$  & (2,1)3 & (0,3)3 & (2,3)3 & (4,1)3  & (2,3)4 
                         & (4,1)4    \\
   & $(\alpha N)\alpha$  & (1,2)3 & (1,4)3 & (1,4)4 &    &    &   \\  \hline

$9/2^+$   
   & $(\alpha \alpha)N$  & (2,2)4 & (0,4)4 & (4,0)4 & (2,4)4 & (4,2)4 
                         &  (4,4)4   \\
   &                     & (2,4)5 & (4,2)5 & (4,4)5  &   &    &    \\
   & $(\alpha N)\alpha$  & (1,3)4 & (1,5)4 & (1,5)5 &    &    &    \\  
\hline\hline

\end{tabular}

\newpage

\begin{table}
\caption{Energies and widths of the unbound states in $^9$Be and $^9$B. 
The energy is from the three-body threshold. The spin and parity of the 
3.065 MeV state of $^9$B is assumed to be $\frac{5}{2}^+$.}
\end{table}

\vspace {2mm}

\begin{tabular}{p{1.5cm}p{1.5cm}p{3.7cm}p{3.7cm}p{2cm}p{2cm}}
\hline\hline

 & & \multicolumn{1}{r}{exp.$^a$}&  &  \multicolumn{1}{r}{cal.} &  \\ \hline

 & $J^{\pi}$ & $E$(MeV$\pm$keV) & ${\it \Gamma}$(MeV$\pm$keV) & 
               $E$(MeV) & ${\it \Gamma}$(MeV) \\ \hline 

 & $3/2^-$ & $-$1.5735  & ------- &  $-$1.431 &  -------  \\

 & $1/2^+$ &  0.111$\pm7$     &  0.217$\pm$10  &           &  \\

 & $5/2^-$  & 0.8559$\pm1.3$   &  0.00077$\pm0.15$ &  0.84  & 0.001  \\

 & $1/2^-$ &  1.21$\pm120$      &  1.080$\pm$110  &  1.20     &  0.46   \\ 

 & $5/2^+$ &  1.476$\pm9$     &  0.282$\pm$11  &  1.98    &  0.6   \\ 

 $^9$Be & $3/2^+$ &  3.131$\pm25$    &  0.743$\pm$55 &  3.3    &  1.6   \\ 

 & $3/2^{-}_{2}$ & 4.02$\pm100^{\:b}$ &  1.33$\pm$360 &   2.9 &  0.8     \\ 

 & $7/2^-$ &  4.81$\pm60^{\:b}$ &  1.21$\pm$230   &  5.03     &  1.2     \\ 

 & $9/2^+$ &  5.19$\pm60^{\:b}$ &  1.33$\pm$90   &  4.9      &  2.9     \\

 & $(1/2^-)$ &  6.37$\pm80$   &  $\sim$1.0  &     &          \\ 

 & $5/2^{-}_{2}$ &      &          &  6.5     &  2.1     \\
\hline

 & $3/2^-$ &  0.277      &  0.00054$\pm0.21$ &  0.30 &  0.004 \\ 

 & $1/2^+$ &  (1.9)        &  $\simeq$0.7    &         &    \\ 

 & $5/2^-$ &  2.638$\pm5$   &  0.081$\pm$5  &  2.55   &  0.044  \\

 & $1/2^-$ &  3.11$^{\:c}$       &  3.1    &  2.73   &  1.0    \\

 & $5/2^+$ &  3.065$\pm30$     & 0.550$\pm$40 &  3.5    &  1.2  \\ 

 $^9$B & $3/2^+$ &     &      &  4.6    &  2.7  \\ 

 & $3/2^-_{2}$ &         &         &  4.2    &  1.4   \\

 & $7/2^-$  &   7.25$\pm60$    &  2.0$\pm$200  &  7.0    &  1.7    \\ 

 & $9/2^+$  &            &       &  6.6    &  3.3    \\

 & $5/2^-_{2}$ &         &       &  8.4    &  2.4    \\
\hline \hline

\end{tabular}

a) Ref.\cite{ajzen}. $\:\:\:$  b) Ref.\cite{dixit}. $\:\:\:$  c) Ref.
\cite{pugh}.

\newpage

\begin{table}
\caption{Radii and electromagnetic properties of $^9$Be. The reduced 
matrix elements are given in Weisskopf units. The bare-nucleon 
charges and $g$-factors are used in the present calculation.  The effective 
charges were used in the shell model calculation of Refs. 
\protect{\cite{glickman}} and \protect{\cite{hees84}} to calculate 
the quadrupole moment and the $E2$ strength. See text for the 
$B(E1)$ value of the present calculation.}
\end{table}

\vspace {0.5mm}

\begin{tabular}{p{1cm}p{3cm}p{3cm}p{1.6cm}p{1.5cm}p{1.5cm}p{1.5cm}p{1.4cm}}
\hline\hline

 $J^{\pi}$ & & exp.$^a$ & present & Ref.\cite{suppl} & Ref.\cite{desc} &
 Ref.\cite{glickman} & Ref.\cite{hees84} \\ \hline \hline

$3/2^-$ & $E$ (MeV)   &  $-$1.5735 & $-$1.431 &   & $-$0.89 &  & \\
  & $r_m$ (fm)   &          &   2.50     & 2.62     &       &  &  \\
  & $r_p$ (fm)   &  2.37$\pm0.01$ &   2.39     &          &       &  &  \\
  & $r_n$ (fm)   &          &   2.58     &          &       &   & \\
  & $\mu$ ($\mu_N$) & $-$1.1778$\pm$0.0009  & $-$1.169 & $-$1.23 &
 $-$1.52 & $-$1.27 & $-$1.070   \\
  & $Q$ (e fm$^2$) & 5.3$\pm$0.3 & 5.13  &  5.76    & 4.77  & 4.35 & 4.66  \\ 
  & $\sigma_R$ (mb)  & 825$\pm$20$^{\:b}$ &  850 &          &    &  &
\\ \hline

 $5/2^-$ & $E$ (MeV)  &  0.8559 & 0.883 &    &  1.89  &  & \\
  & $B(E2;\frac{5}{2}^-$$\rightarrow$$\frac{3}{2}^{-})$  
  & 24.4$\pm$1.8 & 22.0 & 24.7   & 23.5  & 12.5 & $\sim$ 7 \\
  & $B(M1;\frac{5}{2}^-$$\rightarrow$$\frac{3}{2}^{-})$  
  & 0.30$\pm$0.03 & 0.229 &    & 0.10 &  0.23 &   \\ 
\hline 

 $1/2^+$ & $E$ (MeV)  & 0.111  &   &  & 0.05  & & 0.75  \\
  & $B(E1;\frac{1}{2}^+$$\rightarrow$$\frac{3}{2}^{-})$ 
  & 0.22$\pm$0.09 & 0.24  &    & 0.68  & 0.03 & 0.03  \\
  &   &   &  0.18  &   &    &  & \\
\hline 
\hline

\end{tabular}

a) Ref.\cite{ajzen}. $\:\:\:$  b) Ref.\cite{saint}.

\newpage

\vspace {2mm}

\begin{table}
\caption{A set of arrangements and angular momenta included in the 
four-cluster model calculation for the $^9$Be ground state. See Fig. 1(b) 
for the angular 
momenta $\ell_1$, $\ell_2$, and $\ell_3$. The spin of the nucleon 
clusters is coupled to $s_{23}$. The total spin $S$ is restricted to 
$\frac{1}{2}$.}
\end{table}

\vspace {0.5mm}

\begin{tabular}{p{2.5cm}p{3.7cm}p{2.3cm}p{2.3cm}p{3cm}p{1cm}}
\hline\hline

$J^{\pi}$ & arrangement & 
 \multicolumn{3}{l}{angular momentum 
$[(\ell_1,\ell_2)\ell_{12},\ell_3]L$} & $s_{23}$  \\ \hline

$^9$Be: $3/2^-$ 
  & $((t p)\alpha)n$  
  & [(0,0)0,1]1  & [(0,2)2,1]1 & [(0,2)2,1]2 &   1 \\
  & $(t p)(\alpha n)$  
  & [(0,0)0,1]1  & [(0,2)2,1]1 & [(0,2)2,1]2 &   1 \\
  & $((t p)n)\alpha$  
  & [(0,1)1,0]1  & [(0,1)1,2]1 & [(0,1)1,2]2 &   1 \\

  & $((t n)\alpha)p$  
  & [(0,0)0,1]1  & [(0,2)2,1]1 & [(0,2)2,1]2 &   1 \\
  & $(t n)(\alpha p)$  
  & [(0,0)0,1]1  & [(0,2)2,1]1 & [(0,2)2,1]2 &   1 \\
  & $((t n)p)\alpha$  
  & [(0,1)1,0]1  & [(0,1)1,2]1 & [(0,1)1,2]2 &   1 \\

  & $((h n)\alpha)n$  
  & [(0,0)0,1]1  & [(0,2)2,1]1 & [(0,2)2,1]2 &    0 \\
  & $(h n)(\alpha n)$  
  & [(0,0)0,1]1  & [(0,2)2,1]1 & [(0,2)2,1]2 &    0 \\
  & $((h n)n)\alpha$  
  & [(0,1)1,0]1  & [(0,1)1,2]1 & [(0,1)1,2]2 &    0 \\
\hline\hline

\end{tabular}

\newpage
\begin{center}
{\large Figure Captions}
\end{center}

\bigskip

\begin{enumerate}

\item[Fig.\,1.] Different arrangements used in the three-body (a) and 
four-body (b) calculations. The small circles are nucleons, the medium-size 
circle is $\alpha$-particle, and the gray circle is $3N$-cluster, $t$ or 
$h$. The orbital angular momenta for the relative motion between the 
clusters connected by solid lines are denoted by $\ell_i$. The spin of 
the clusters is $s_i=\frac{1}{2}$; the spin of the $\alpha$-particle is zero 
and it is omitted.

\item[Fig.\,2.] Complex eigenvalues for $J^{\pi}=\frac{3}{2}^-$ (a) and 
$\frac{7}{2}^-$ (b) of $^9$Be. The rotation angle $\theta$ is in units of 
radian. The point indicated by an open circle corresponds to a resonance. 

\item[Fig.\,3.] Experimental and calculated energies of $^9$Be (a) and 
$^9$B (b) from the three-body threshold. The data are from Refs. 
\cite{glickman,dixit,ajzen}. The 3.065 MeV state of $^9$B is assumed to be 
$\frac{5}{2}^+$. 

\item[Fig.\,4.] Elastic charge form factor for $^9$Be. The data are from 
Refs. \cite{meyer,bern,slight}.

\item[Fig.\,5.] Elastic transverse form factor for $^9$Be. The data are 
from Refs. \cite{rand,lapikas}. 

\item[Fig.\,6.] Monopole (a) and quadrupole (b) density distributions of 
protons and neutrons for the ground state of $^9$Be. 

\item[Fig.\,7.] Monopole density distributions of 
protons and neutrons, (a) in linear scale 
and (b) in logarithmic scale, for the excited $\frac{1}{2}^+$ state and the 
$\frac{3}{2}^-$ ground state of $^9$Be. The value of $u=1.0$ is used 
for the $\frac{1}{2}^+$ state.  

\item[Fig.\,8.] Spectroscopic amplitudes of the ground state of $^9$Be 
for the $^8$Be+$n$ arrangement. The symbols $r$ and $R$ denote the 
distances of two $\alpha$-particles and of $n$ from their 
center-of-mass. The set of angular momenta is (a) $\ell_1=0, 
\ell_2=1, L=1$, (b) $\ell_1=2, \ell_2=1, L=1$, and 
(c) $\ell_1=2, \ell_2=1, L=2$.

\item[Fig.\,9.] Spectroscopic amplitudes of the ground state of $^9$Be for 
the $^5$He+$\alpha$ arrangement. The symbol $r$ is the 
distance between $n$ and $\alpha$ and $R$ the distance between 
their center-of-mass and $\alpha$. The set of angular momenta is (a) 
$\ell_1=1, \ell_2=0, L=1$, (b) $\ell_1=1, \ell_2=2, L=1$, and 
(c) $\ell_1=1, \ell_2=2, L=2$.

\end{enumerate}

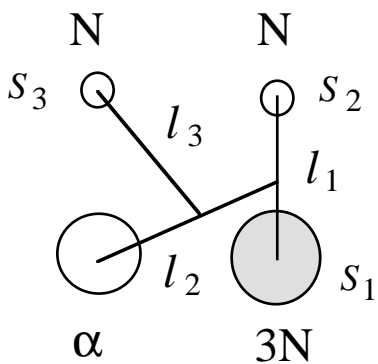
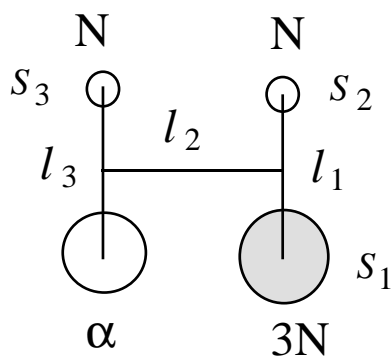
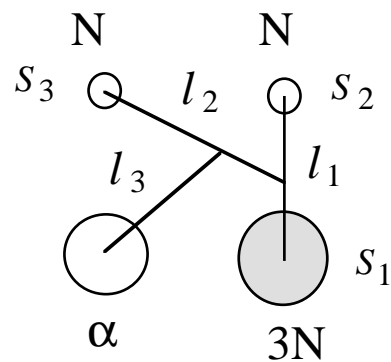
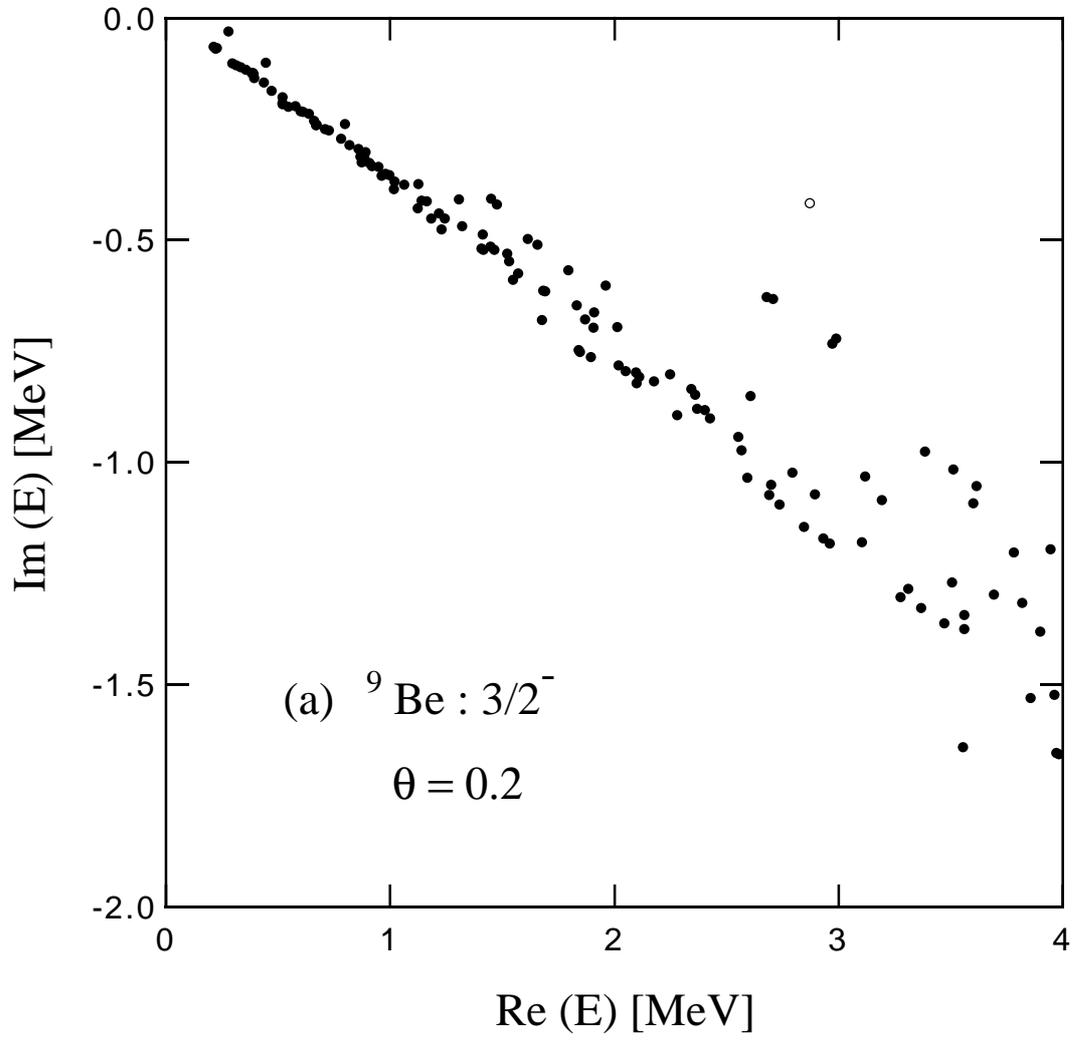
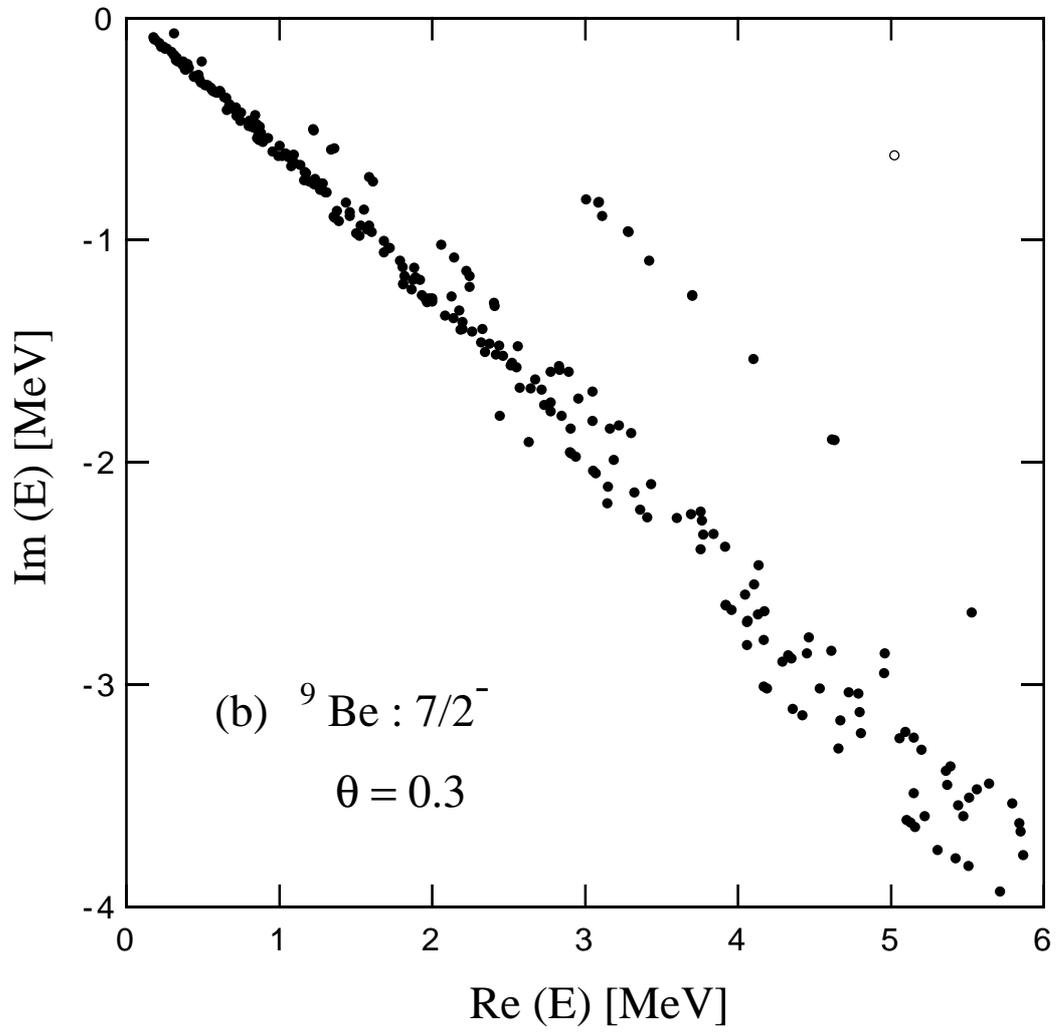
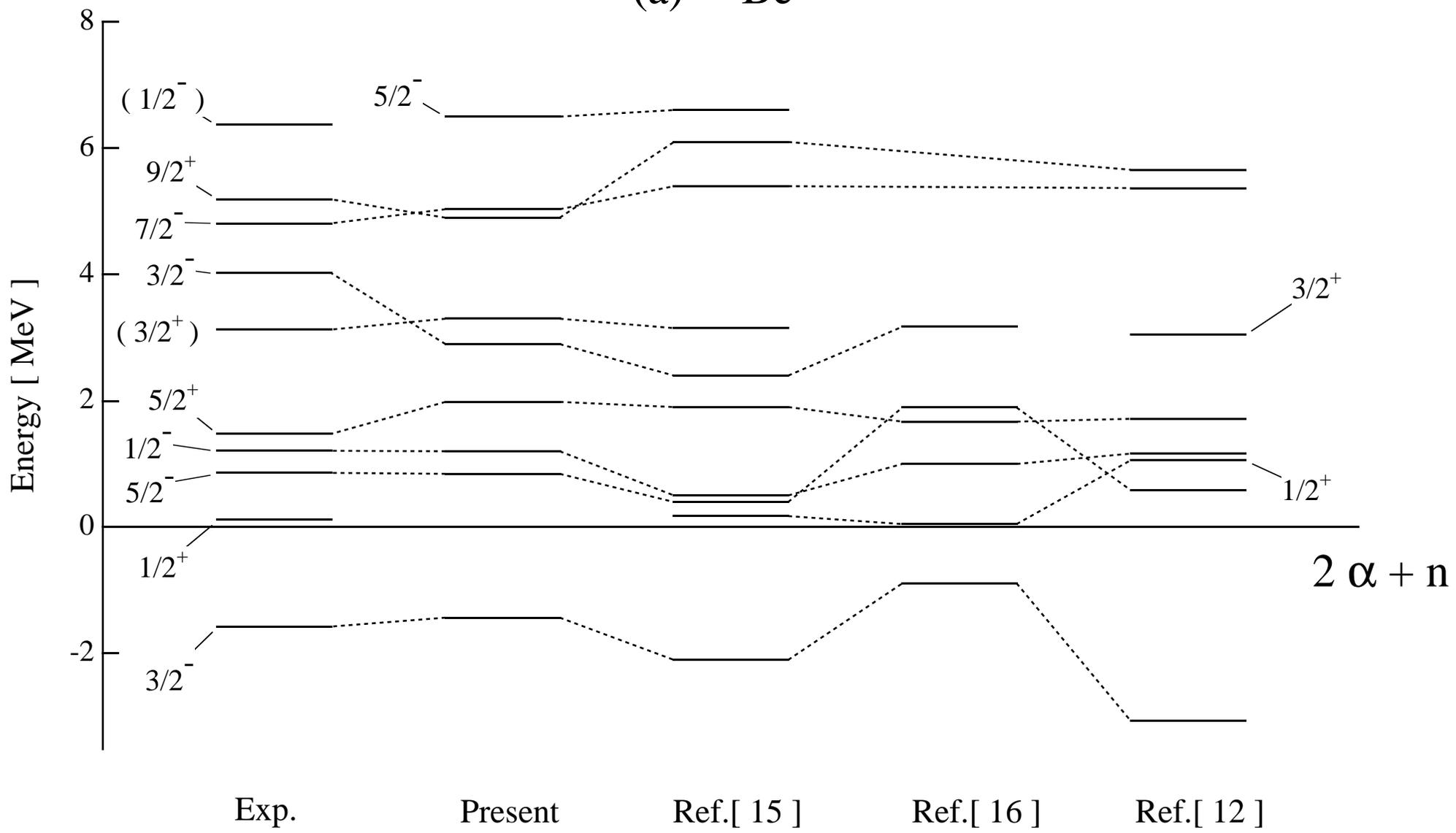
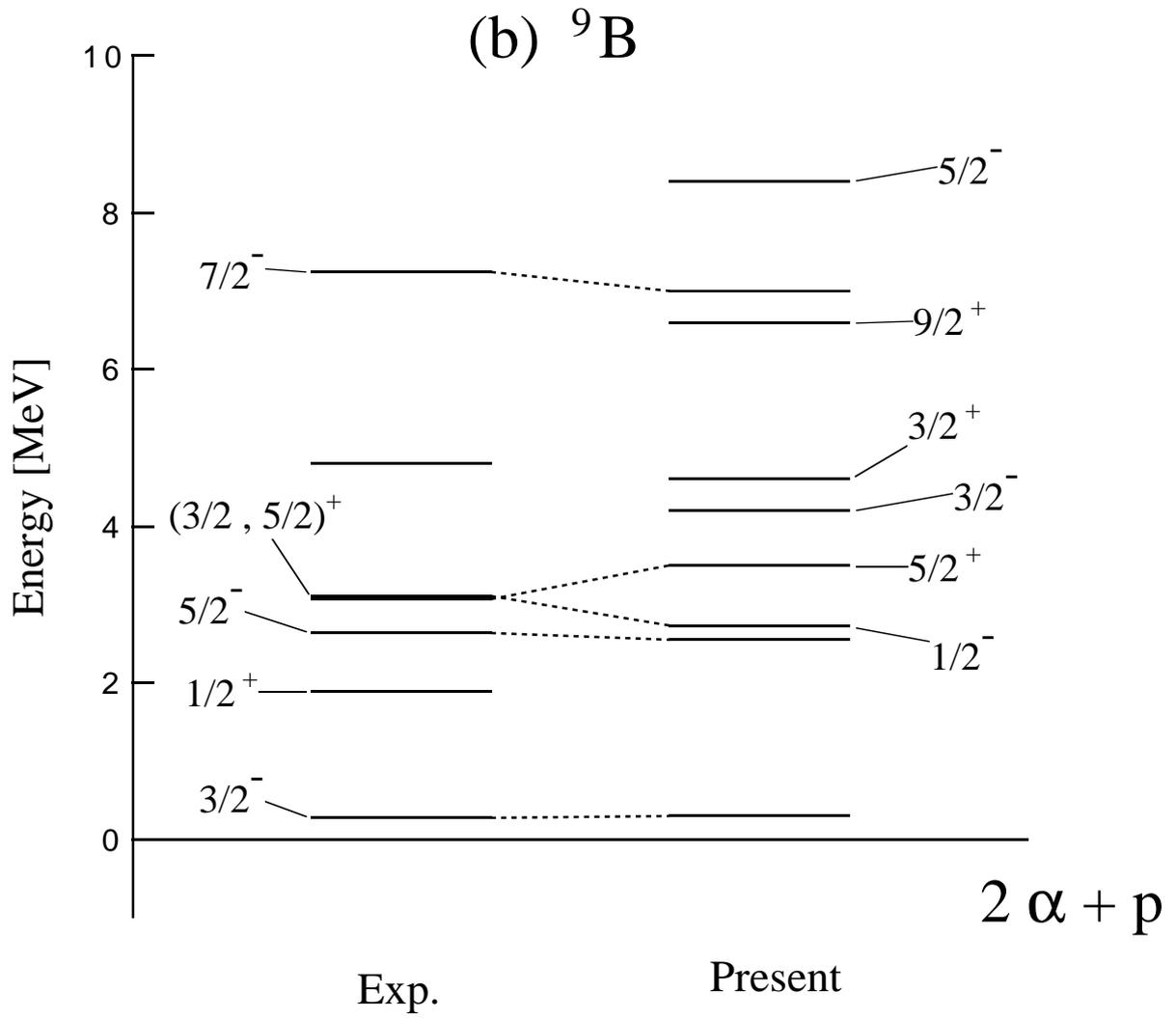
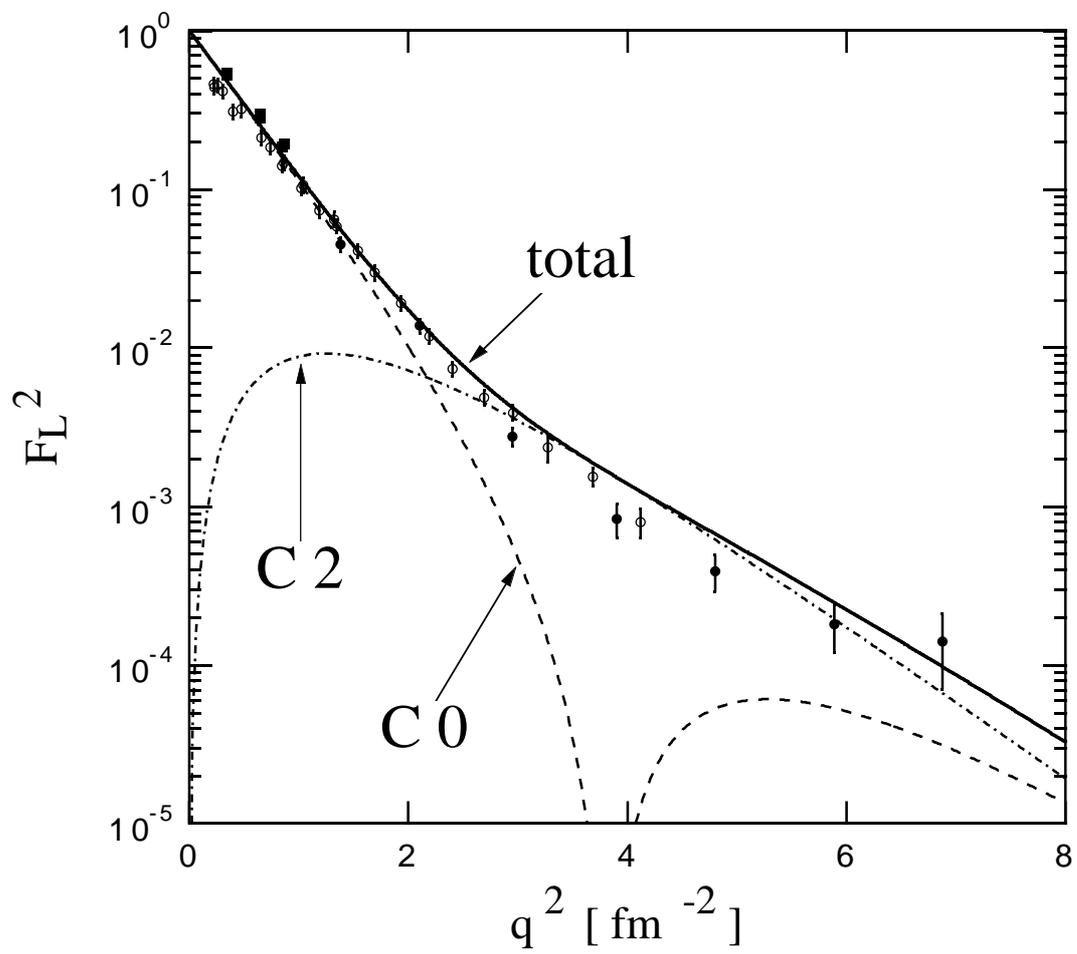
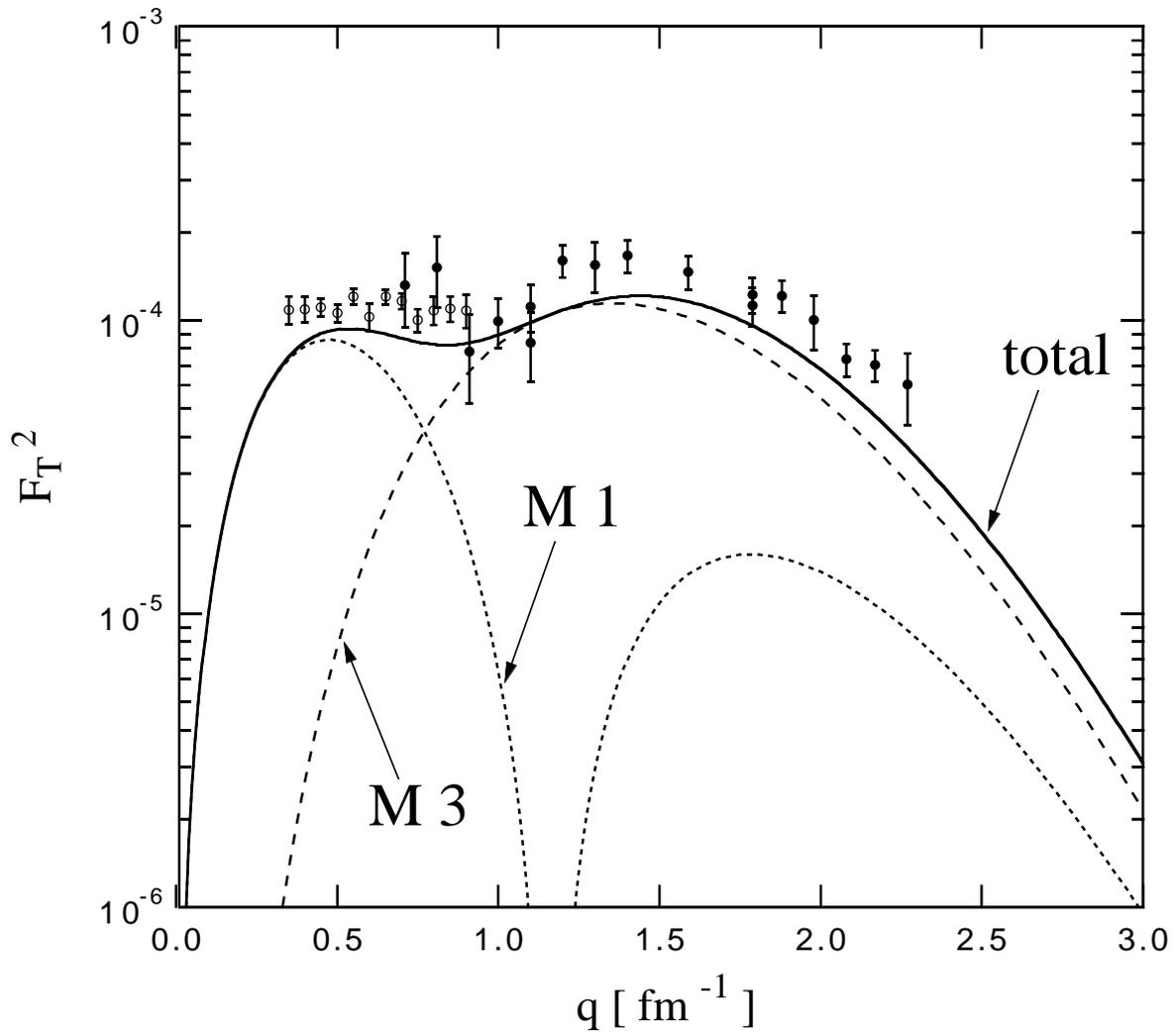
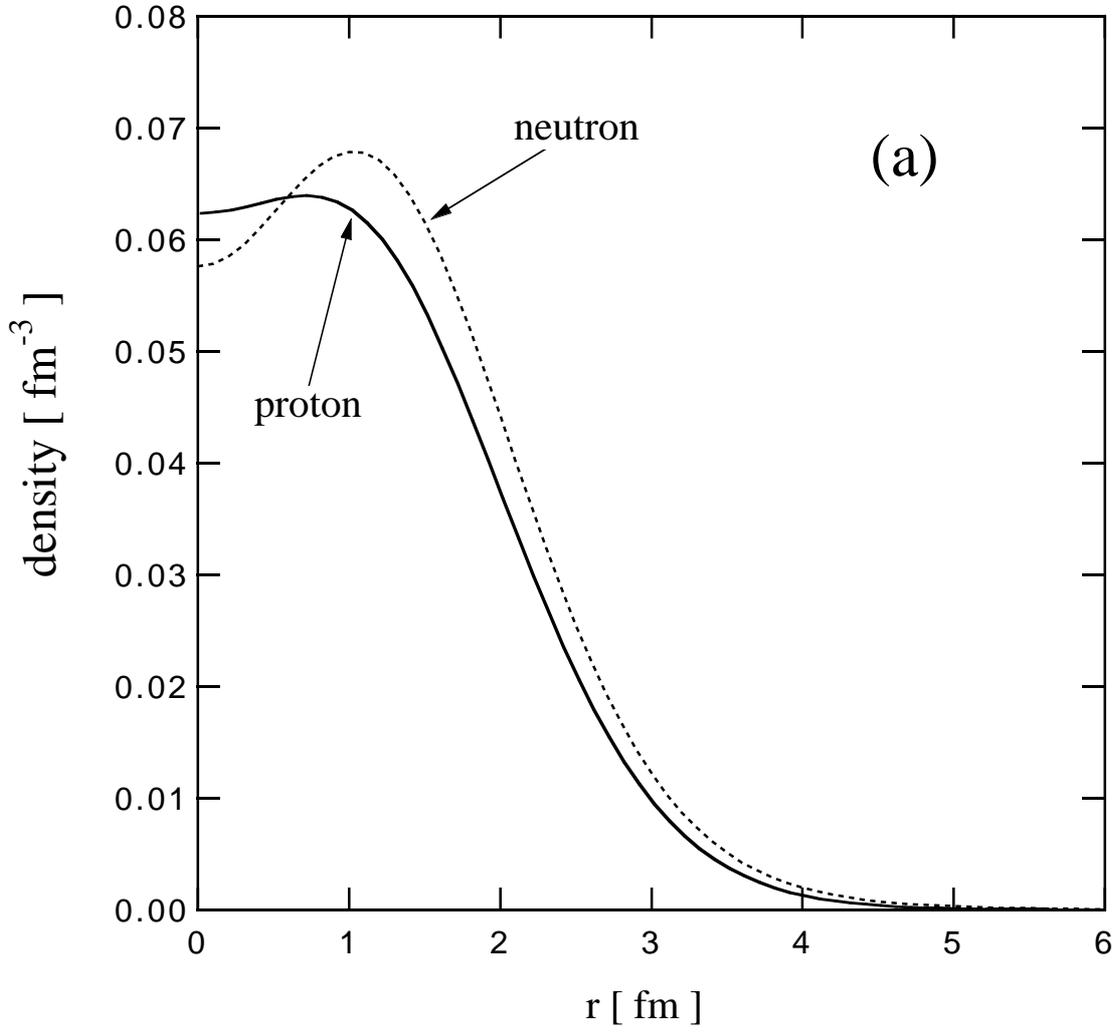
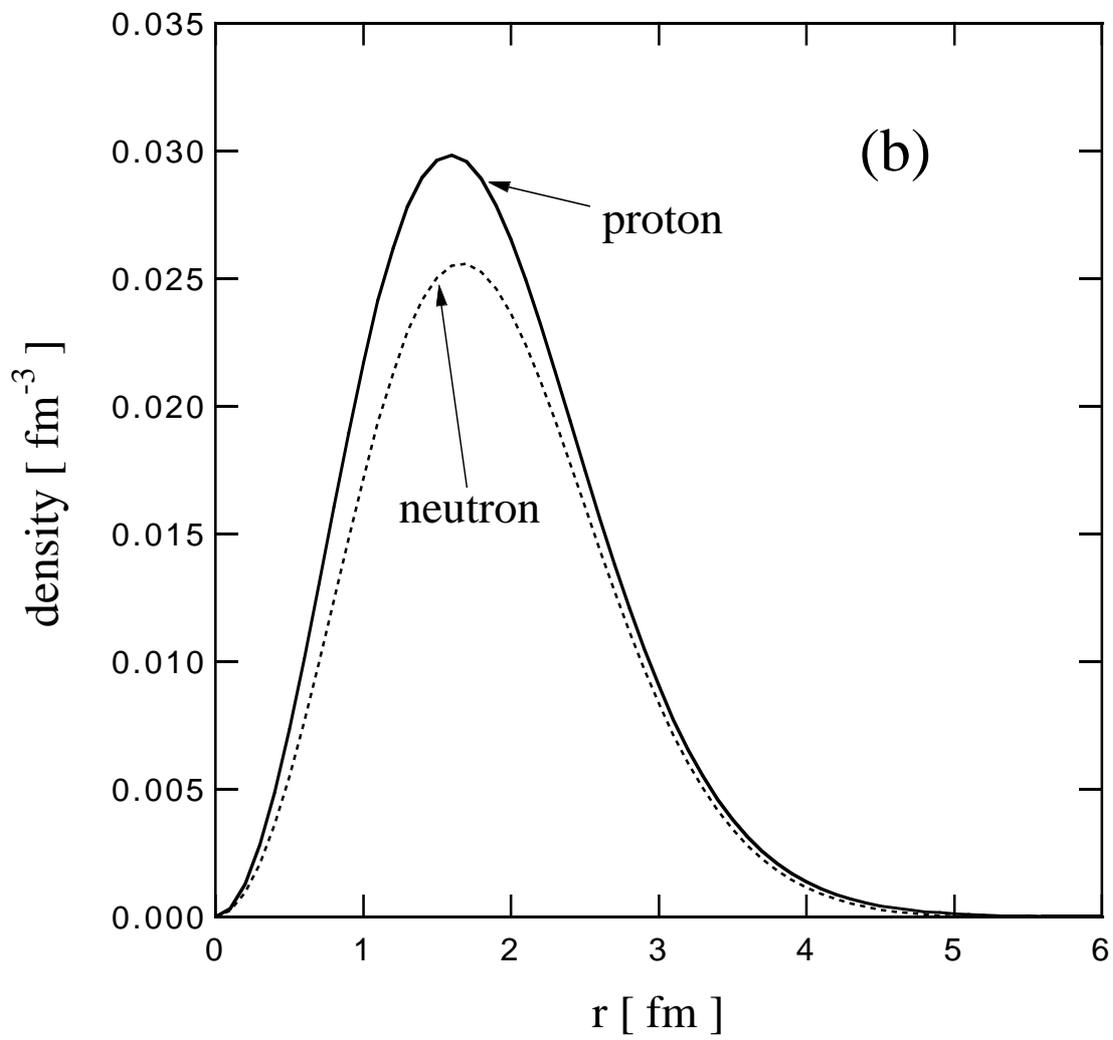
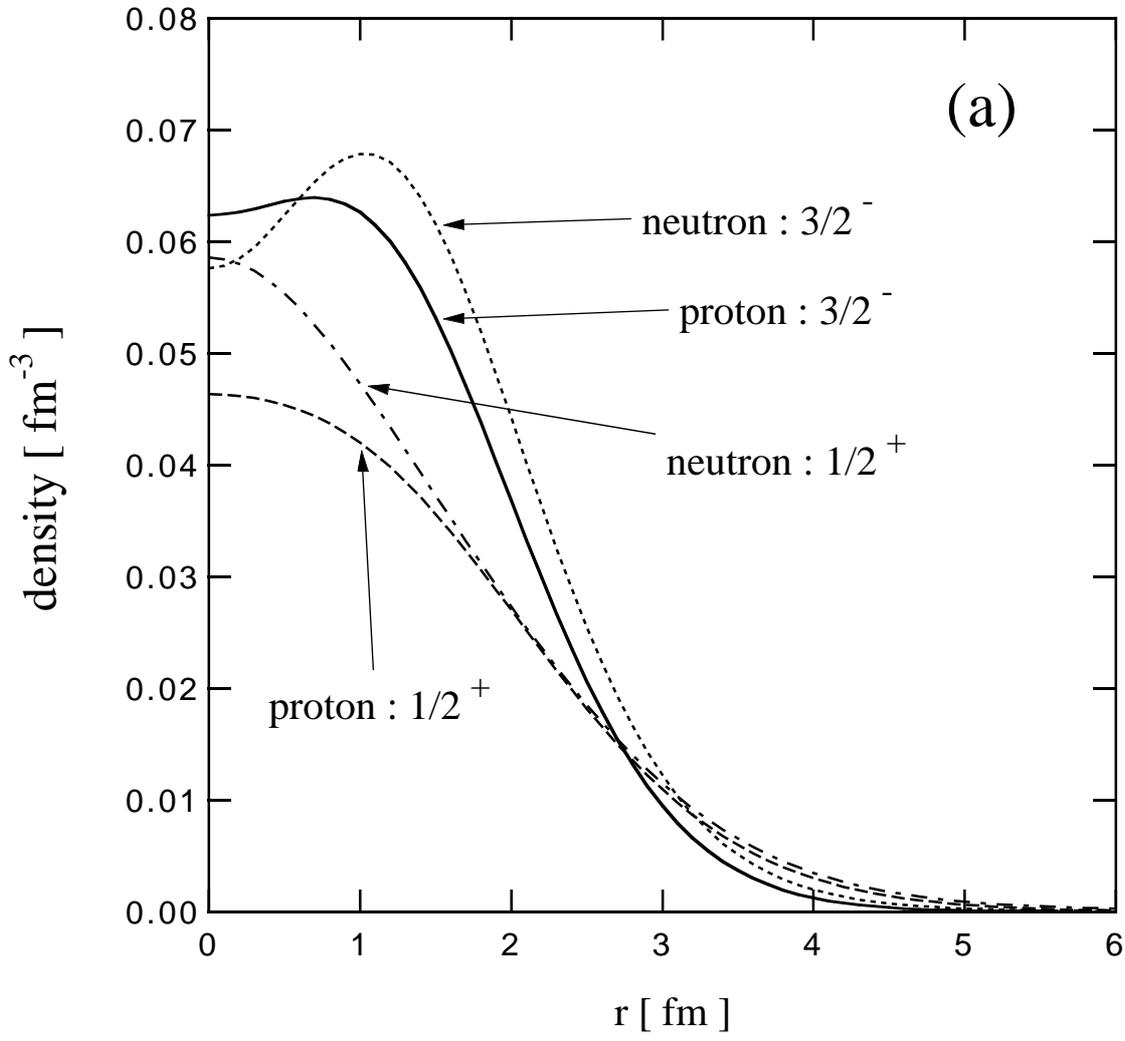
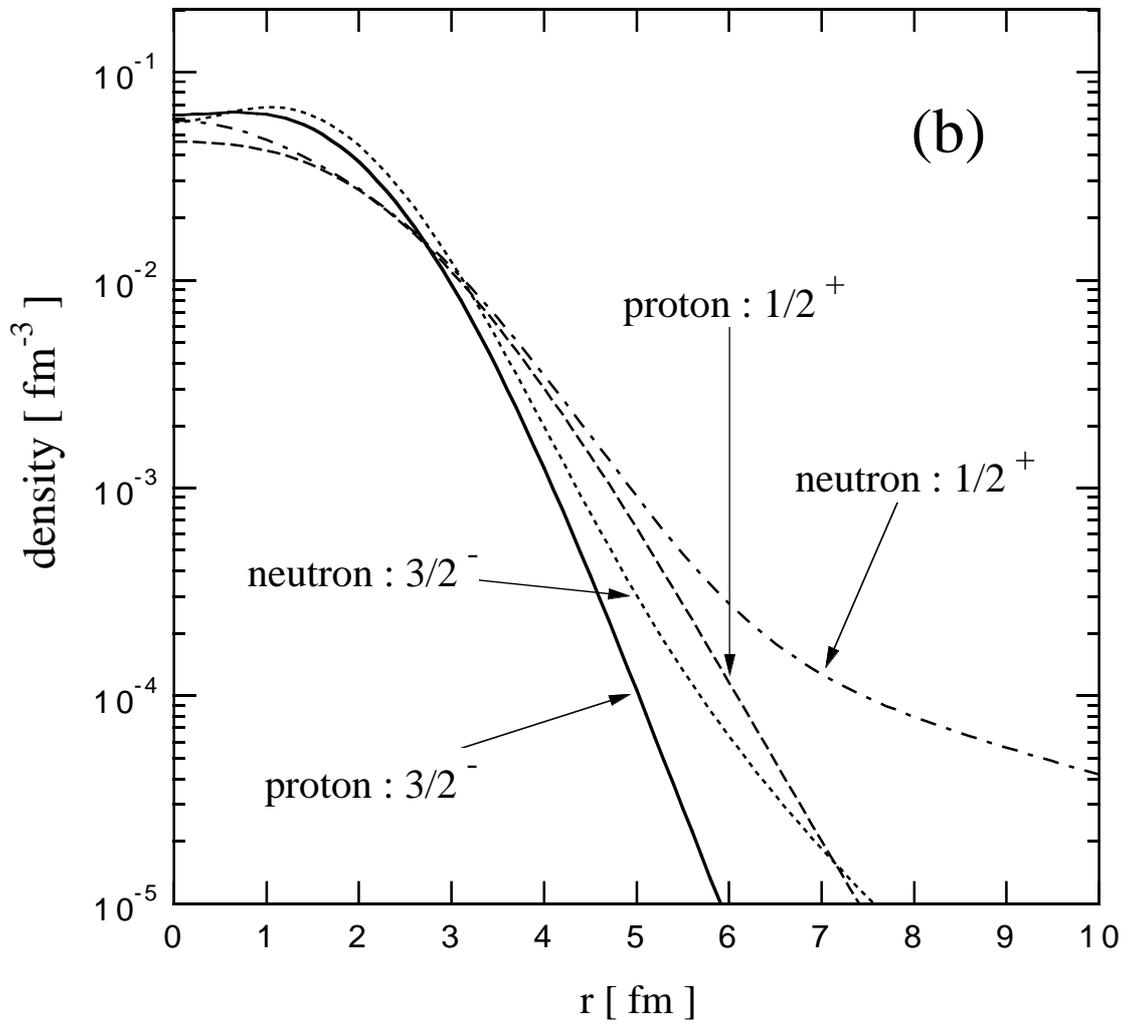
\end{document}